\documentclass[authoryear,preprint,review,8pt]{elsarticle}

\usepackage{natbib}
\setcitestyle{numbers,square}
\usepackage{amssymb}
\usepackage{bm}
\usepackage{amsmath}
\usepackage[lined,boxed,commentsnumbered,ruled]{algorithm2e}
\usepackage{siunitx}
\usepackage{subfigure}
\usepackage{graphicx} 
\usepackage{epstopdf}
\usepackage{booktabs}
\usepackage{xcolor}
\usepackage{changes}
\usepackage{todonotes}
\usepackage{hyperref}

\usepackage{multirow}
\usepackage{booktabs}
\usepackage{array}
\usepackage{amsmath}
\usepackage{siunitx}
\usepackage{xcolor}
\colorlet{red}{black}

\journal{Journal of Computational Physics}

\begin{document}

\begin{frontmatter}



\title{A Hybrid Gas-Kinetic Scheme and Discrete Velocity Method for Continuum and Rarefied Flows}

 \author[label1]{Hangkong Wu}
 \author[label1]{Yuze Zhu}
 \author[label3]{Yajun Zhu}
 \author[label1,label4,label5]{Kun Xu\corref{cor1}}
\cortext[cor1]{Corresponding author}
 \ead{makxu@ust.hk}
\affiliation[label1]{organization={Department of Mathematics, Hong Kong University of Science and Technology},
             city={ Clear Water Bay, Kowloon},
             state={Hong Kong},
             country={China}}
 \affiliation[label3]{organization={Research and Development Office, Shanghai Suochen Information Technology Co., Ltd},
             city={Shanghai},
             country={China}}
 \affiliation[label4]{organization={Department of Mechanical and Aerospace Engineering, Hong Kong University of Science and Technology},
             city={ Clear Water Bay, Kowloon},
             state={Hong Kong},
             country={China}}
\affiliation[label5]{organization={Shenzhen Research Institute, Hong Kong University of Science and Technology},
             city={Shenzhen},
             country={China}}

\begin{abstract}

The gas‑kinetic scheme (GKS) provides high computational efficiency and accuracy for continuum flow simulations but is unable to reliably capture rarefaction effects. In contrast, although the discrete velocity method (DVM) is better suited for rarefied flows, it exhibits reduced accuracy and slow convergence when applied to continuum regimes. To overcome these limitations, this work proposes a hybrid GKS–DVM method that integrates the strengths of both approaches. The hybrid approach balances the equilibrium distribution function in GKS with the upwind‑reconstructed non-equilibrium distribution function in DVM through a numerical collision time. This balancing strategy ensures to recover Navier-Stokes solutions in the continuum limit (asymptotic preserving), while naturally capturing free molecular flows in the rarefied limit. Moreover, the introduction of a numerical collision time significantly enhances robustness in shock capturing for continuum flow applications. To further reduce computational cost of the hybrid approach, several adaptive strategies based on the prescribed Knudsen number and the maximum Mach number have been proposed. The effectiveness and accuracy of the proposed hybrid method are systematically assessed through four representative test cases: a flat‑plate boundary layer, a lid‑driven cavity flow, shock structures, and flow past a semi-cylinder. The first case is subjected to continuum conditions, while the remaining cases span a broad range of Knudsen numbers. The results demonstrate that the proposed method achieves high solution accuracy and computational efficiency across both continuum and rarefied flow regimes.

\end{abstract}


\end{frontmatter}

\section{Introduction}\label{introduction}

Space vehicles operating from near‑ground altitudes to the near‑space environment encounter multiscale flow regimes spanning from continuum to rarefied conditions~\cite{Blanchard1997}. In the rarefied regime, the continuum assumption underlying the Navier–Stokes (NS)~\cite{ROE1997250} equations breaks down, rendering conventional NS-based computational fluid dynamics methods, like the gas-kinetic scheme (GKS)~\cite{XU19949, XU2001289}, inaccurate. Boltzmann-based solution methods, such as the discrete velocity method (DVM)~\cite{WANG201833, MIEUSSENS2000}, are capable of analyzing rarefied flows; however, they generally suffer from low computational efficiency and slow convergence in continuum flow regimes due to the stiffness associated with the collision term~\cite{YUAN2020106972}. Furthermore, the upwind reconstruction used for flux evaluation at cell interfaces may introduce excessive numerical dissipation in the continuum limit, resulting in reduced solution accuracy.

To address such multiscale flows, the unified gas‑kinetic scheme (UGKS) was proposed in 2010 ~\cite{XU20107747, Huang_Xu_Yu_2012, Huang_Xu_Yu_2013}. UGKS employs the integral solution of the Boltzmann equation based on the Bhatnagar–Gross–Krook (BGK)~\cite{BGK1954} or the BGK-Shakhov models~\cite{Shakhov1968, Xu8145382} to construct the time‑dependent gas distribution function at the cell interface. The resulting distribution function couples particle transport and collision effects and is subsequently used to evaluate numerical fluxes. Owing to the coexistence of equilibrium and non‑equilibrium components in the distribution function, UGKS naturally recovers the Navier–Stokes solutions in the continuum regime~\cite{LIU2016305, LIUModified, CHEN201552} while accurately capturing rarefaction effects in free‑molecular flows~\cite{LIU201496}. Therefore, UGKS constitutes a multiscale approach applicable across all Knudsen number regimes. Over the past fifteen years, UGKS has been widely adopted to analyze a broad spectrum of flow problems. To enhance its computational efficiency, various extensions of UGKS~\cite{Guo2021}, including implicit~\cite{ZHU2019190, ZHU201616}, adaptive~\cite{Long2024, CHEN20126643}, multi-grid~\cite{Zhu2017} and memory-saving~\cite{ZHANG2025109684} formulations, have been developed. In addition, the discrete unified gas-kinetic scheme (DUGKS) provides another unified kinetic framework for multiscale gas flow simulations~\cite{Guo2013, Shan2020}.

However, UGKS and DUGKS are slightly computationally intensive due to the evaluation of the time-dependent distribution function at the cell interface.
Over the last decade, several efficient hybrid and multiscale approaches have been proposed~\cite{Chen2016, ZHANG2025108367}. 
Kolobov et al.~\cite{KOLOBOV2007589} employed domain decomposition algorithms to separate rarefied and continuum flow regions. In the rarefied regime, a direct solution of the Boltzmann equation is used, while kinetic schemes are applied in the continuum regime. The results demonstrate the high computational efficiency of the unified flow solver. 
Wu et al.~\cite{ZHU2021110091, WuLei2020} proposed the general synthetic iterative scheme (GSIS) to accelerate the solution of the Boltzmann equation in multiscale gas flows, especially in near‑continuum regimes.
Yang et al.~\cite{Yang2018, YANG2019738} developed an improved DVM that blends the Navier–Stokes (NS) flux with the DVM flux using the physical collision time as a weighting parameter. In the continuum limit, the NS flux becomes dominant, whereas in the rarefied limit, the scheme reduces to the conventional DVM, thereby exhibiting the asymptotic‑preserving property. However, the incorporation of the NS flux requires an explicit discretization of the Navier–Stokes equations, which increases the difficulty of solver development.
To circumvent this limitation, Yuan et al.~\cite{YUAN2021105473, Zhang2023} proposed a multiscale DVM in which the interface flux is decomposed into equilibrium and non‑equilibrium components. The non‑equilibrium flux is evaluated in the same manner as in the conventional DVM, while the equilibrium flux is further split into an Euler flux and a kinetic flux vector splitting (KFVS) flux to enhance numerical stability. These two contributions are balanced through a numerical collision time and the physical local time step (similar to DUGKS). A series of benchmark tests have demonstrated the effectiveness of this multiscale approach across different flow regimes.

Unlike the above work, an easy-to-implement, efficient and simultaneously accurate multiscale hybrid approach is proposed in this work. As is well known, GKS provides an efficient and accurate framework for continuum‑flow simulations, whereas the DVM is well suited for rarefied flows. Motivated by these complementary strengths, this work proposes a hybrid approach that combines GKS and DVM for applications spanning both continuum and rarefied flow regimes. The proposed method hybridizes the equilibrium distribution function in GKS with the non‑equilibrium distribution function reconstructed via upwind discretization in DVM, with their respective contributions balanced through a numerical collision time.
The introduction of the numerical collision time enables the hybrid scheme to recover the Chapman–Enskog expansion in the continuum limit for smooth flows. In continuum flows with shock waves, the contribution of the non‑equilibrium DVM flux is increased, providing additional numerical dissipation and thereby ensuring stable shock‑capturing capability. In the rarefied limit, the hybrid approach naturally recovers free‑molecular flow behavior. The advantage of the newly proposed approach over the aforementioned hybrid approach is that the numerical flux is directly evaluated from the distribution function, without relying on any extra discretization of the macroscopic governing equations.
Furthermore, several adaptive strategies based on the Knudsen number and Mach number are introduced.
\begin{enumerate}[\qquad $\boldsymbol \cdot$]
\item In the continuum and subsonic regime, only the equilibrium GKS flux is evaluated.

\item In the continuum but transonic or supersonic regime, both equilibrium and non‑equilibrium flux contributions are considered.

\item In all other flow regimes, only the DVM flux is employed.
\end{enumerate}
Through these adaptive strategies, the proposed hybrid approach is expected to achieve improved effectiveness across both continuum and rarefied flow regimes.

The remainder of this paper is organized as follows. Section~\ref{methodology} introduces the numerical methods adopted in this work, including the newly proposed hybrid approach. Section~\ref{boundary_condition} describes the boundary conditions employed in this work. Section~\ref{test_case} presents the results of four benchmark test cases, namely the flat‑plate boundary layer, two‑dimensional lid‑driven cavity flows, one‑dimensional shock structures, and flow past a semi-cylinder. Finally, concluding remarks are given in Section~\ref{conclusion}.

\section{Methodology}\label{methodology}
In this section, the BGK model is first introduced, followed by the detailed introduction of the DVM, GKS, and the newly proposed hybrid approach.

\subsection{BGK Model}

The gas-kinetic equation based on the BGK model is given by
\begin{equation}\label{BGK}
f_t + \boldsymbol u \cdot \nabla_{\boldsymbol x} f = \frac{g-f}{\tau}
\end{equation}
where $f$ is the distribution function, $\boldsymbol u$ is the particle velocity vector, $\boldsymbol x$ is the grid coordinate, $\tau=\mu/p$ is the molecular mean collision time related to the physical viscosity, $\mu$ is the dynamic viscosity, $p$ is the static pressure, and $g$ is the equilibrium distribution function satisfying the following Maxwellian distribution, given by
\begin{equation}\label{Max_g}
g=\Big ( \frac{\lambda}{\pi} \Big)^{\frac{K+2}{2}}e^{-\lambda [(\boldsymbol u - \boldsymbol U)^2 + {\boldsymbol \xi}^2]}
\end{equation}
where $K$ is the number of internal degree of freedom and is equal to 3 for a diatomic gas (including two rotational degrees-of-freedom and one transitional degree-of-freedom in the z direction), $\lambda=\rho/2p$ with $\rho$ being the density, $\boldsymbol U$ represents the macroscopic flow velocity vector, and $\boldsymbol \xi^2=\sum_{i=1}^{K}\boldsymbol \xi_i^2$ is the internal variable.

The dynamic viscosity can be calculated from the variable hard-sphere model,
\begin{equation}
\mu = \mu_{ref}\Big( \frac{T}{T_{ref}}\Big)^{\omega}
\end{equation}
where $T_{ref}$ is the reference temperature, $\omega$ is the index related to the variable hard-sphere model and equal to 0.81, $\mu_{ref}$ is the reference viscosity and its definition is given by
\begin{equation}\label{Kn}
Kn =\frac{\lambda_{mfp}}{L} =\frac{4\alpha (5-2\omega)(7-2\omega)}{5(\alpha + 1)(\alpha + 2)}\sqrt{\frac{m_0}{2\pi k_B T}}\frac{\mu}{\rho L}
\end{equation}
where $\lambda_{mfp}$ is the mean free path, $\alpha = 1$ and $\omega = 0.5$ for the hard-sphere model, $m_0$ is the molecule mass, $k_B$ is the Boltzmann constant ($k_B=R_g\cdot m_0$ and $R_g$ denotes the gas constant), $L$ is the reference length of the computational domain, $Kn$ is the Knudsen number. The relationship between $Kn$, Mach number ($Ma$) and Reynolds number ($Re$) is given by
\begin{equation}
Kn\cdot Re = \frac{4\alpha(5-2\omega)(7-2\omega)}{5(\alpha +1)(\alpha + 2)}\frac{\gamma}{2\pi}Ma
\end{equation}
where $\gamma$ is the specific heat ratio.

The collision term in Eq.~\ref{BGK} meets the requirement of the compatibility condition
\begin{equation}
\int(g-f)\boldsymbol \psi d\boldsymbol \xi du = 0
\end{equation}
where $\boldsymbol \psi=[1,\boldsymbol u,1/2(\boldsymbol u^2 + \boldsymbol \xi^2)]^T$ is the collision invariants.

The connection between the gas distribution function and the macroscopic flow variables and fluxes are given by
\begin{gather}
\boldsymbol Q = \int_{\boldsymbol u}\int_{\boldsymbol \xi} f \boldsymbol \psi d\boldsymbol \xi d\boldsymbol u\label{macro_q}\\
\boldsymbol F^{\boldsymbol Q} = \int_t \int_{\boldsymbol u}\int_{\boldsymbol \xi} f\boldsymbol \psi \boldsymbol u d\boldsymbol \xi d\boldsymbol u dt
\label{macro_f}
\end{gather}

\subsection{Discrete Velocity Method}

If an upwind scheme is employed for spatial discretization and the trapezoidal rule is used to approximate the collision term, Eq.~\ref{BGK} can be rewritten in the following discretized finite‑volume form.
\begin{equation}\label{dBGK}
f_{i,j}^{k+1} = f_{i,j}^{k} - \frac{1}{\Delta V}\int_{0}^{\Delta t}\int_{\partial V}\boldsymbol n\cdot \boldsymbol u_j f_{i+1/2,j}^{DVM}(t) dS dt + \frac{\Delta t}{2} \Big(\frac{g_{i,j}^{k}-f_{i,j}^k}{\tau^k}+ \frac{g_{i,j}^{k+1}-f_{i,j}^{k+1}}{\tau^{k+1}}\Big)
\end{equation}
Where $\Delta V$ and $\partial V$ represent the volume and surface of the control volume respectively, $\boldsymbol n$ represents the unit outward normal vector on the surface of the control volume, the subscripts $i$ and $i+1/2$ denote the variables at the cell center and interface respectively, $j$ represents the discrete velocity point, $k$ and $k+1$ denote the time marching level, $\Delta t$ is the time interval restrained by the following Courant-Friedrichs-Lewy (CFL) condition
\begin{equation}
\Delta t = \sigma \frac{\Delta x}{max(\xi_{max},U_{max}) + c}
\end{equation}
where $\xi_{max}$ is the maximum discrete particle velocity, $U_{max}$ is the maximum macroscopic flow velocity,
and $c$ represents the speed of sound.

Making some arrangements to Eq.~\ref{dBGK} leads to
\begin{equation}\label{dBGK2}
f_{i,j}^{k+1} =A\Big[ f_{i,j}^{k} - \frac{1}{\Delta V}\int_{\partial V}\boldsymbol n\cdot \boldsymbol u_j F_{i+1/2,j}^{DVM} dS  + \frac{\Delta t}{2} \Big(\frac{g_{i,j}^{k}-f_{i,j}^k}{\tau^k}+ \frac{g_{i,j}^{k+1}}{\tau^{k+1}}\Big) \Big]
\end{equation}
where $A=\Big(1+\frac{\Delta t}{2\tau^{k+1}}\Big)^{-1}$, and $F_{i+1/2,j}^{DVM}$ is the time integral solution of the distribution function in DVM, defined by
\begin{equation}\label{Ff}
F_{i+1/2,j}^{DVM} = \int_{0}^{\Delta t}f_{i+1/2,j}^{DVM}(t)dt
\end{equation}

If first-order time accuracy is adopted, the time-evolution distribution function $f_{i+1/2,j}^{DVM}(t)$ in Eq.~\ref{Ff} can be expressed by
\begin{equation}\label{ft}
f_{i+1/2,j}^{DVM}(t) = f_{0,i+1/2,j}
\end{equation}
where $f_{0,i+1/2,j}$ is the initial distribution function at the cell interface.

$f_{0,i+1/2,j}$ can be reconstructed using an upwind scheme, as shown in the following equation
\begin{gather}
\begin{aligned}
f_{0,i+1/2,j} = f_{0,i,j} +  &\beta_i \nabla_{\boldsymbol x}f_{i,j} \cdot (\boldsymbol x_{i+1/2} - \boldsymbol x_{i})H[u_{n,j}] + \\
&\beta_{i+1}\nabla_{\boldsymbol x}f_{i+1,j} \cdot (\boldsymbol x_{i+1} - \boldsymbol x_{i+1/2})(1-H[u_{n,j}])
\end{aligned}
\end{gather}
In the above equation,
$\nabla_{\boldsymbol x}f_{i,j}$ and $\nabla_{\boldsymbol x}f_{i+1,j}$ are the spatial gradients of the distribution function, which are evaluated using the least-square method in this work. The coefficients
$\beta_i$ and $\beta_{i+1}$ are the scaling factors determined by the limiter~\cite{VENKATAKRISHNAN1995120, Ji2021AGC, Zhang2022AHC}. $u_{n,j}$ is the normal part of the j-th discrete particle velocity, and $H[u_{n,j}]$ is the Heaviside function, defined as
\begin{equation}
H[u_{n,j}]=
\begin{cases}
1, & u_{n,j}\geq 0 \\
0, & u_{n,j}\textless 0
\end{cases}
\end{equation}

Substituting Eq.~\ref{ft} into Eq.~\ref{dBGK2} yields the time integral solution of the distribution function in DVM at the cell interface
\begin{gather}\label{fintf}
F_{i+1/2,j}^{\textit{DVM}} = \int_{0}^{\Delta t} f_{0,i+1/2,j}dt
=f_{0,i+1,j}\Delta t
\end{gather}

Substituting Eq.~\ref{fintf} into Eq.~\ref{macro_f} leads to the macroscopic flux at the interface
\begin{gather}\label{Fq_int}
\begin{aligned}
\boldsymbol F_{i+1/2}^{\boldsymbol Q, DVM} &= \int_{\boldsymbol u} \int_{\boldsymbol \xi}F_{i+1/2,j}^{DVM}\boldsymbol \psi \boldsymbol u \cdot \boldsymbol n d\boldsymbol ud\boldsymbol \xi = \int_{\boldsymbol u} \int_{\boldsymbol \xi} f_{0,i+1,j}\Delta t\boldsymbol u \cdot \boldsymbol n \boldsymbol \psi d\boldsymbol ud\boldsymbol \xi\\
& = \sum_{j} w_j \Big(\int_{\boldsymbol \xi} f_{0,i+1,j}\boldsymbol u_j \cdot \boldsymbol n \boldsymbol \psi d\boldsymbol \xi \Big) \Delta t
\end{aligned}
\end{gather}

Based on the above equation, the macroscopic flow variables can be updated using the following finite-volume approach
\begin{equation}\label{Q_update}
\boldsymbol Q^{k+1} = \boldsymbol Q^{k} -\frac{1}{\Delta V}\int_{\partial V}  \boldsymbol F_{i+1/2}^{\boldsymbol Q, DVM} dS
\end{equation}

The updated $\boldsymbol Q^{k+1}$ can be used to compute $g^{k+1}$ in Eq.~\ref{dBGK2} using Eq.~\ref{Max_g} and also the collision time $\tau^{k+1}$. All variables in Eq.~\ref{dBGK2} have been computed, and thereby the distribution function at the cell interface can be updated.

\textit{Remark 1. } To analyze flows with an arbitrary Prandtl number, the BGK-Shakhov model ($f^+$) will be used to replace the BGK model ($g$), which is given by
\begin{equation}
f^{+} = g\Big[1+(1-Pr)\boldsymbol c\cdot \boldsymbol q\Big (\frac{c^2}{RT}-5 \Big)/(5pRT)\Big]=g+g^{+}
\end{equation}

\textit{Remark 2.} To remove the dependence of the numerical integration on the internal variable in Eq.~\ref{Fq_int}, the reduced function $h$ and $b$ are introduced, defined as
\begin{equation}
h = \int_{\boldsymbol \xi} f d\boldsymbol \xi, b = \int_{\boldsymbol \xi}f \boldsymbol \xi^2 d\boldsymbol \xi
\end{equation}

\subsection{Gas-Kinetic Scheme}

Unlike DVM which directly solves the Boltzmann equation numerically, GKS uses the integral solution of the distribution function to reconstruct the distribution function at the cell interface, given by
\begin{gather}\label{int_BGK}
f^{GKS}(\boldsymbol x_{i+1/2},\boldsymbol u_j,\boldsymbol \xi,t) = \frac{1}{\tau}\int_{0}^{t}g(\boldsymbol x^{'},\boldsymbol u_j,\boldsymbol \xi,t^{'})e^{-(t-t^{'})/\tau}dt^{'} +e^{-t/\tau}f_0(\boldsymbol x_0,\boldsymbol u_j,\boldsymbol \xi,0)
\end{gather}
where $\boldsymbol x^{'}= \boldsymbol x_{i+1/2} - \boldsymbol u_j (t-t^{'})$ is the particle trajectory, $\boldsymbol x_0=\boldsymbol x_{i+1/2} - \boldsymbol u_j t$ is the initial position of a particle, $f_0$ is the initial distribution function. To simplify notation, we set $\boldsymbol x_{i+1/2}=0$ at the cell interface.

According to the Chapman-Enskog expansion, $f_0$ in Eq.~\ref{int_BGK} can be expressed by
\begin{gather}\label{glr}
\begin{aligned}
f_0(\boldsymbol x_0,\boldsymbol u_j,\boldsymbol \xi, 0) =& g^{L}(\boldsymbol 0, \boldsymbol u, \boldsymbol \xi, 0)[1-(t+\tau)\boldsymbol a^{L}\cdot \boldsymbol u_j-\tau A^{l}]H[u_{n,j}] +\\
&  g^{R}(\boldsymbol 0, \boldsymbol u,\boldsymbol \xi, 0)[1-(t+\tau)\boldsymbol a^{R}\cdot \boldsymbol u_j-\tau A^{r}](1-H[u_{n,j}])
\end{aligned}
\end{gather}

According to the Taylor expansion, $g$ in Eq.~\ref{int_BGK} can be expressed by
\begin{equation}\label{gc}
g(\boldsymbol x^{'},\boldsymbol u_j,\boldsymbol \xi,t^{'}) = g^C(\boldsymbol 0,\boldsymbol u_j,\boldsymbol \xi,0)(1-\boldsymbol a^c\cdot \boldsymbol u_jt^{'}+A^ct^{'})
\end{equation}

In Eqs.~\ref{glr} and ~\ref{gc}, $g^L$ and $g^R$ are the equilibrium gas distribution at the left and right hand sides of an interface and are determined by the reconstructed macroscopic flow variables $Q^L$ and $Q^R$, and $g^C$ is the equilibrium distribution function computed by the kinetic-averaged macroscopic flow variables $\boldsymbol Q^C$ at the cell interface. $\boldsymbol a^L$, $\boldsymbol a^R$ and $\boldsymbol a^C$ are the spatial derivatives of $g^L$, $g^R$ and $g^C$, while $A^L$, $A^R$ and $A^C$ are the temporal derivatives of $g^L$, $g^R$ and $g^C$. Their definitions are given by
\begin{gather}
\boldsymbol a^{L/R/C}=\frac{1}{g^{L/R/C}}\frac{\partial g^{L/R/C}}{\partial\boldsymbol x} = \frac{ \partial ln g^{L/R/C}}{\partial\boldsymbol x}\\
A^{L/R/C}=\frac{1}{g^{L/R/C}}\frac{\partial g^{L/R/C}}{\partial t} = \frac{ \partial ln g^{L/R/C}}{\partial t}
\end{gather}

Substituting Eqs.~\ref{glr} and ~\ref{gc} into Eq.~\ref{int_BGK}, the distribution function at the cell interface can be expressed by
\begin{gather}\label{df}
\begin{aligned}
f^{GKS}(\boldsymbol x_{i+1/2},\boldsymbol u_j,\boldsymbol \xi,t)=&(1-e^{-\frac{t}{\tau_n}})g^C+\\
&[(t+\tau)e^{-\frac{t}{\tau_n}}-\tau]\boldsymbol a^C\cdot \boldsymbol u_jg^C +\\
&(t-\tau+\tau e^{-\frac{t}{\tau_n}})A^C g^C + \\
&e^{-\frac{t}{\tau_n}}[g^L H[u_{n,j}] + g^R(1-H[u_{n,j}])]-\\
&e^{-\frac{t}{\tau_n}}(\tau + t)[\boldsymbol a^L\cdot \boldsymbol u_j g^L H[u_{n,j}] + \boldsymbol a^R\cdot \boldsymbol u_j g^R (1-H[u_{n,j}])] - \\
&\tau e^{-\frac{t}{\tau_n}}[g^LA^L H[u_{n,j}] + g^RA^R(1-H[u_{n,j}])]
\end{aligned}
\end{gather}
where $\tau_n$ is the numerical collision time, defined by
\begin{equation}
\tau_n = \frac{\mu}{p} + C\Big |\frac{p_L - p_R}{p_L + p_R}\Big |\Delta t
\end{equation}
where $p_L$ and $p_R$ are the static pressure at the left and right hand side of an interface, and $C$ is a constant ranging from 1 to 10.

Substituting Eq.~\ref{df} into Eq.~\ref{macro_f} yields the macroscopic flux at the cell interface, which is subsequently used to update the macroscopic flow variables via Eq.~\ref{Q_update}.
Since the distribution function $f$ in Eq.~\ref{df} can be connected to the equilibrium distribution function, the macroscopic flux at the interface can be obtained directly by taking velocity moments of $f$, without discretizing the velocity space.

\textit{Remark 3.} In the continuum and subsonic flow regimes, where both $\Delta t\gg \tau$ and $\Delta t\gg \tau_n$ are satisfied, the integral solution of the BGK equation will be reduced to
\begin{equation}\label{gks_f_simp2}
f_{i+1/2,j}^{GKS}(t)=
f^{GKS}(\boldsymbol x_{i+1/2},\boldsymbol u_j,\boldsymbol \xi,t)=g^C\Big(1-\tau(\boldsymbol a^C\cdot \boldsymbol u_j +A^c) +A^C t\Big)
\end{equation}

Its time integral solution is given by
\begin{gather}
\begin{aligned}
F_{i+1/2,j}^{GKS}(t)&=
\int_{0}^{\Delta t} f_{i+1/2,j}^{GKS}(t)dt\\
&=\int_{0}^{\Delta t}g^C\Big(1-\tau(\boldsymbol a^C\cdot \boldsymbol u_j +A^c)+ A^C  t\Big)dt\\
&=g^C\Big[1-\tau(\boldsymbol a^C\cdot \boldsymbol u_j +A^c)\Big]\Delta t +\frac{1}{2} A^C \Delta t^2
\end{aligned}
\end{gather}

\subsection{Hybrid GKS and DVM}

In rarefied flow regimes, the non‑equilibrium distribution function deviates significantly from its equilibrium counterpart; as a result, approximating $f_0$ using the Chapman–Enskog expansion in the GKS framework can introduce substantial errors. In contrast, DVM reconstructs the distribution function at the cell interface solely through an upwind scheme, without accounting for the equilibrium contribution, which leads to excessive numerical dissipation in continuum and subsonic flow regimes. Although this deficiency can be alleviated by mesh refinement, it comes at the expense of significantly increased computational cost. Moreover, the stiff collision term in the continuum regime results in slow convergence of DVM. Consequently, GKS is generally unsuitable for rarefied-flow simulations, while DVM encounters notable difficulties in accurately and efficiently resolving continuum flows.

To overcome these limitations, a hybrid approach that couples the GKS with the DVM through a numerical collision time is proposed to enable accurate simulations across both continuum and rarefied flow regimes. This hybrid framework leverages the complementary strengths of the two methods. Specifically, the equilibrium component of the distribution function in the GKS, as given in Eq.~\ref{gks_f_simp2}, is combined with the upwind‑reconstructed distribution function in the DVM, as defined in Eq.~\ref{ft}, yielding the following time‑evolution form of the distribution function:
\begin{equation}\label{fhybrid}
f_{i+1/2,j}^{hybrid}(t) = (1-e^{-\frac{\Delta t}{\tau_n}}) f_{i+1/2,j}^{GKS}(t) + e^{-\frac{\Delta t}{\tau_n}}f_{i+1/2,j}^{DVM}(t)
\end{equation}

Based on asymptotic‑preserving analysis, it can be shown that in the continuum and subsonic limit, where $\Delta t \gg \tau_n$ and $e^{-\Delta t/\tau_n} \to 0$, the hybrid time‑evolution distribution function reduces to the equilibrium distribution in the GKS, thereby recovering the Navier–Stokes solutions. In the continuum transonic and supersonic regimes, as the ratio $\Delta t/\tau_n$ increases and $e^{-\Delta t/\tau_n}$ becomes non‑negligible, the DVM‑type distribution function dominates in shock regions. This enhances numerical dissipation and improves the robustness of the scheme for shock capturing. In rarefied flow regimes, where $\Delta t \ll \tau_n$, the hybrid time‑evolution distribution function naturally recovers the free‑molecular flow limit.

To further improve computational efficiency of the hybrid approach based on the above analysis, several adaptive strategies based on the flow feature, Mach number and Knudsen number are proposed in this work.

\begin{enumerate}[\quad $\cdot$]
\item For steady flow analysis, $A^{C}t$ in GKS, as shown in Eq.~\ref{gks_f_simp2}, will diminish, leading to the following steady distribution function at the cell interface
\begin{gather}\label{fhybrid2}
\begin{aligned}
f_{i+1/2,j}^{hybrid} &= (1-e^{-\frac{\Delta t}{\tau_n}})f_{i+1/2,j}^{GKS} + e^{-\frac{\Delta t}{\tau_n}}f_{i+1/2,j}^{DVM}\\
&=(1-e^{-\frac{\Delta t}{\tau_n}})g^C[1-\tau (\boldsymbol a^C\cdot \boldsymbol u_j+ A^C)]+e^{-\frac{\Delta t}{\tau_n}}f_{0,i+1/2,j}
\end{aligned}
\end{gather}


\item For continuum and subsonic steady flows where $Ma<1$ and $Kn\leq0.001$, $\Delta t\gg \tau_n$ and $e^{-\frac{\Delta t}{\tau_n}} \to 0$, the hybrid distribution function at the interface will be further reduced to
\begin{equation}\label{fgks}
f_{i+1/2,j}^{hybrid} = f_{i+1/2,j}^{GKS} = g^C[1-\tau (\boldsymbol a^C\cdot \boldsymbol u_j + A^C)]
\end{equation}

\item In the free‑molecular ($Kn \ge 10$), transition ($10^{-1} \le Kn < 10$), and slip ($10^{-2} \le Kn < 10^{-1}$) flow regimes, the solution is dominated by the DVM distribution function. Consequently, the hybrid distribution function can be reduced to
\begin{equation}\label{fdvm}
f_{i+1/2,j}^{hybrid}= f_{i+1/2,j}^{DVM} = f_{0,i+1/2,j}
\end{equation}

\end{enumerate}

According to the above analysis, the distribution function of the hybrid approach at the cell interface for steady flow field analysis can be written as follows

\begin{equation}\label{hybrid_f}
f_{i+1/2,j}^{hybrid}=
\begin{cases}
f^{GKS}_{i+1/2,j}, & \text{$Ma<1$ and $Kn \leq Kn_c$} \\
(1-e^{-\frac{\Delta t}{\tau_n}})f^{GKS}_{i+1/2,j} + e^{-\frac{\Delta t}{\tau_n}}f^{DVM}_{i+1/2,j}, & \text{$Ma\geq 1$ and $Kn \leq Kn_c$}\\
f^{DVM}_{i+1/2,j}, & Kn>Kn_c
\end{cases}
\end{equation}

In the above equation, $Kn_c$ denotes the critical Knudsen number, which is set to 0.001 in this study based on numerical testing.

\section{Solid Wall Boundary Condition}\label{boundary_condition}

The hybrid approach requires the treatment of boundary conditions for both the equilibrium distribution function in the GKS and the upwind‑reconstructed distribution function in the DVM. The former can be conveniently imposed using ghost cells. However, the implementation of solid wall boundary conditions for the latter is more challenging. In this work, the diffuse reflection boundary condition is adopted to model isothermal walls, while the no‑slip bounce‑back boundary condition is employed to treat adiabatic no‑slip walls. For slip walls, a symmetric boundary condition is applied. To save space, the detailed derivations are omitted here; interested readers are referred to Ref.~\cite{YUAN2021105473} for further details.

\section{Test Cases}\label{test_case}

Four benchmark cases are employed to verify the solution accuracy and computational efficiency of the hybrid approach. The first case is the flat‑plate boundary layer in the continuum flow regime. The remaining cases involve lid‑driven cavity flows, shock structures, and flow past a semi-cylinder, respectively, all covering a wide range of Knudsen numbers.

\subsection{Case 1: Flat-Plate Boundary Layer}

\begin{figure}[h!]
    \centering
\includegraphics[width=0.9\linewidth]{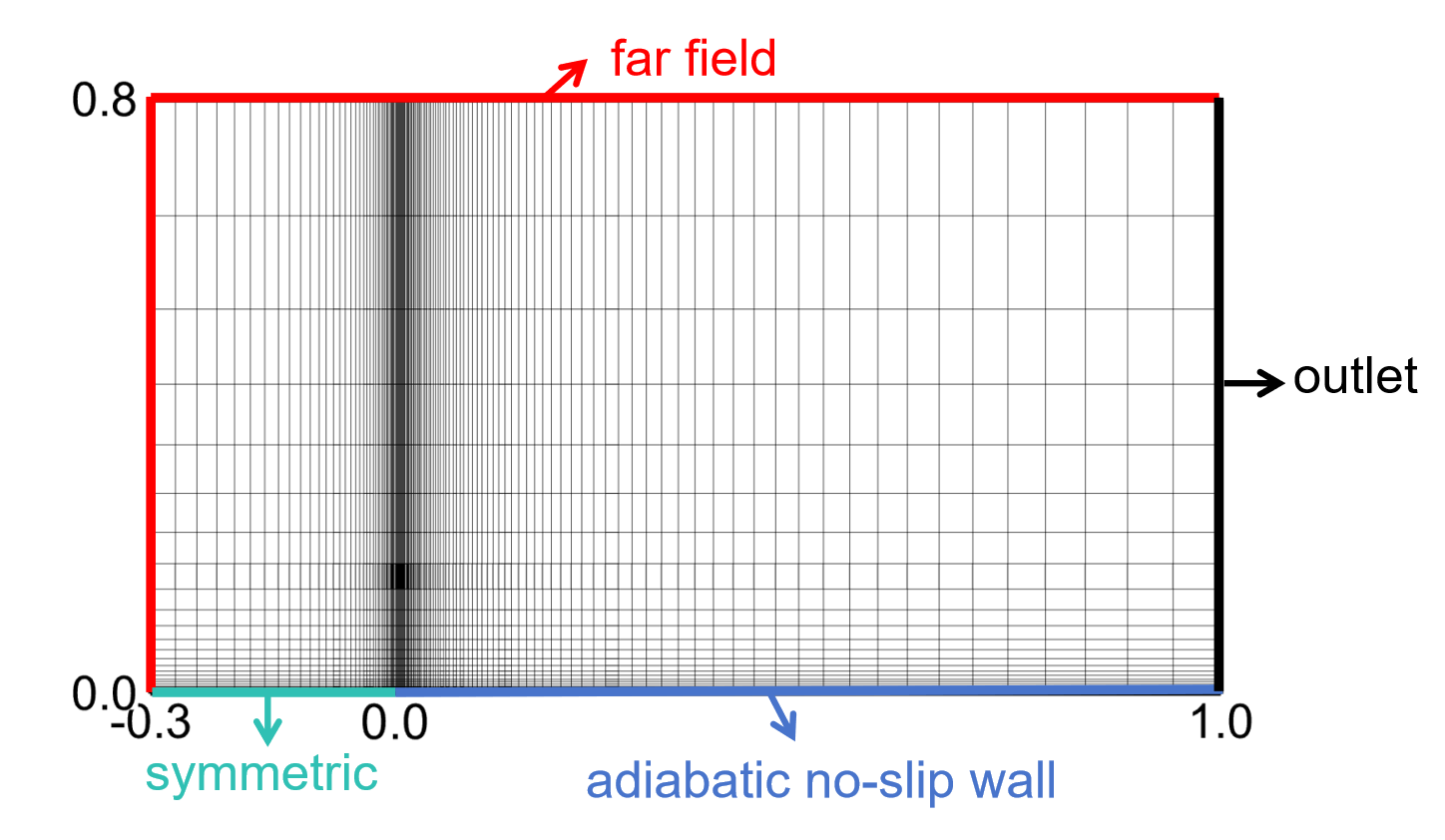}
    \caption{The computational mesh of the flat plate boundary layer}
   \label{bl_mesh}
\end{figure}

Figure~\ref{bl_mesh} shows the computational mesh for the flat-plate boundary layer. The number of grid points is 121 including 80 on the flat plate in the x direction and 36 in the y direction. The cell size close to the wall is equal to $1.75\times 10^{-4}$.
The freestream Mach number is set to 0.15, the freestream Reynolds number is set to $1.0\times10^5$, and the Prandtl number is 1.0.

\begin{figure}[h!]
\centering
\subfigure[U]{
	\includegraphics[width=2.25in]{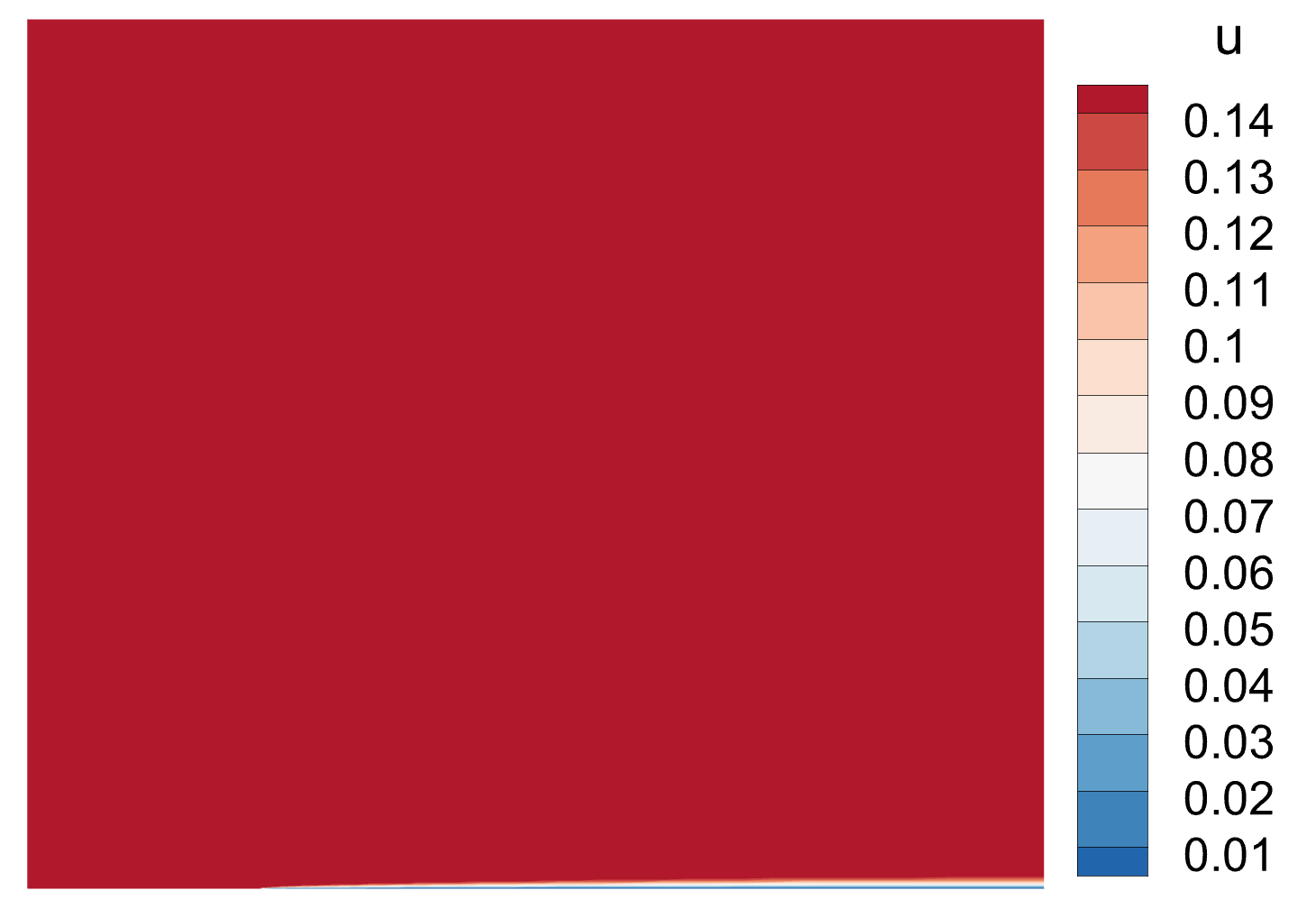}
 \label{plate_U0}
 }
 \subfigure[V]{
	\includegraphics[width=2.25in]{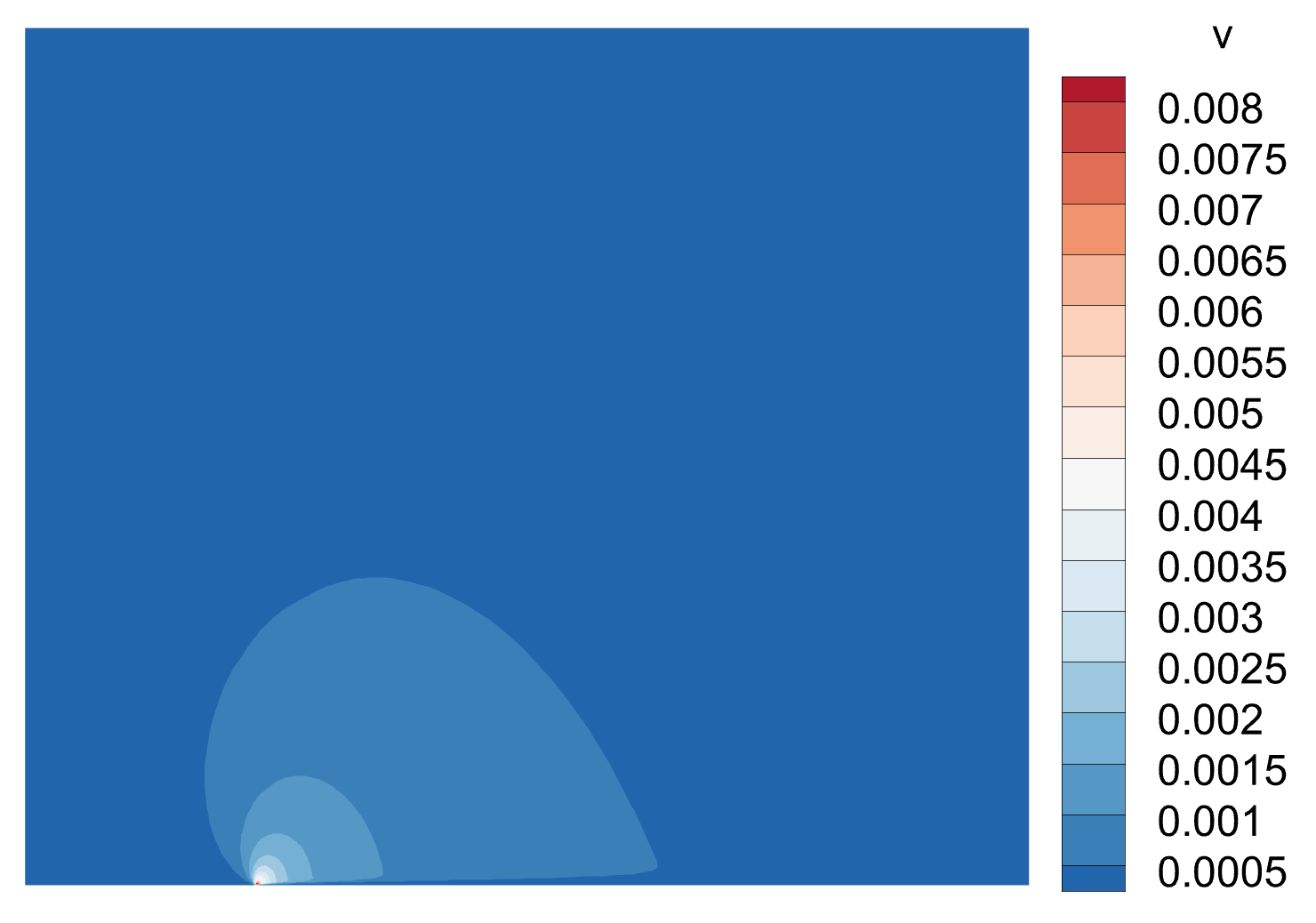}
 \label{plate_V0}
 }
  \subfigure[hybrid]{
	\includegraphics[width=2.25in]{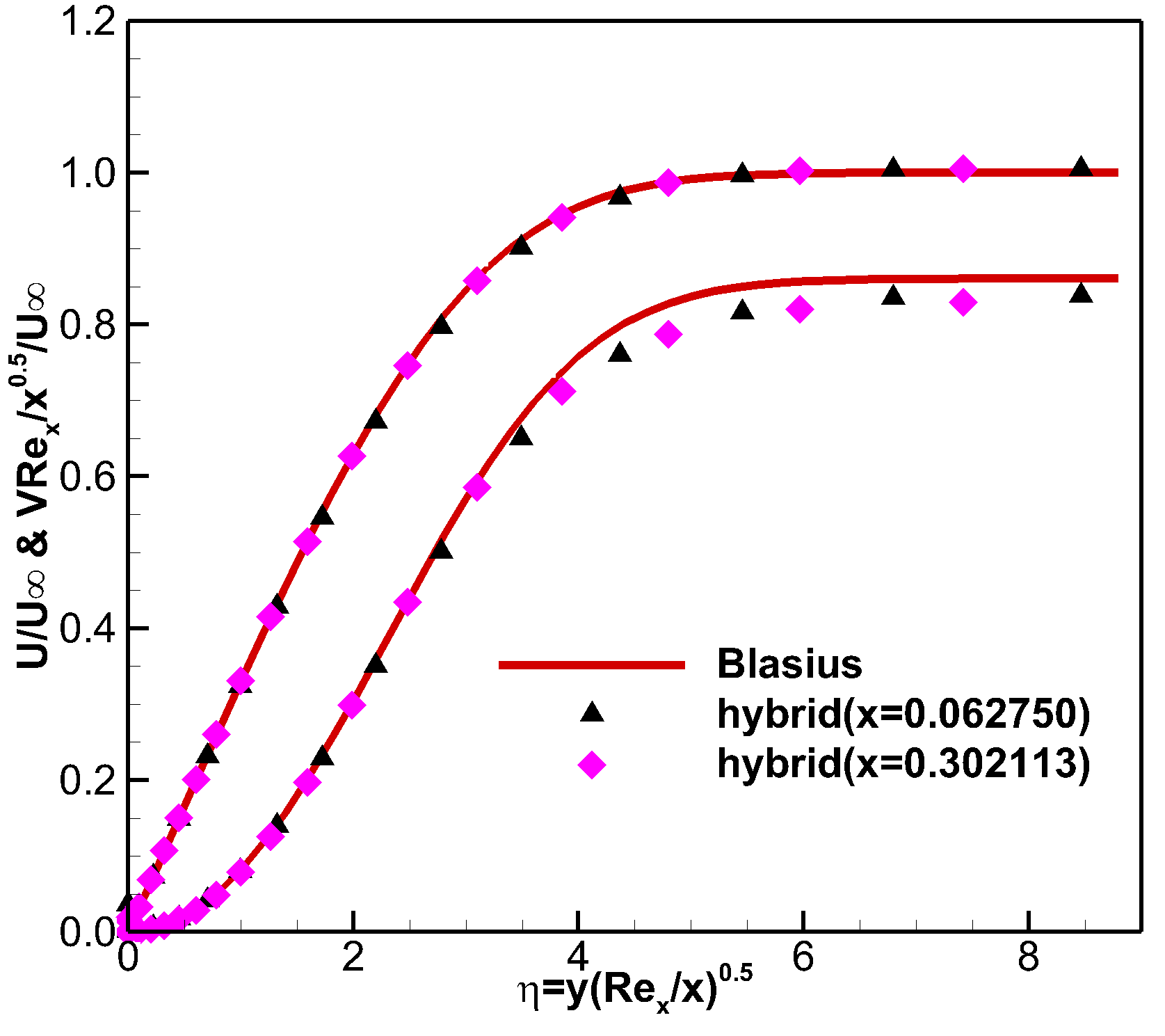}
 \label{plate_uv_hybrid}
 }
 \subfigure[adaptive]{
	\includegraphics[width=2.25in]{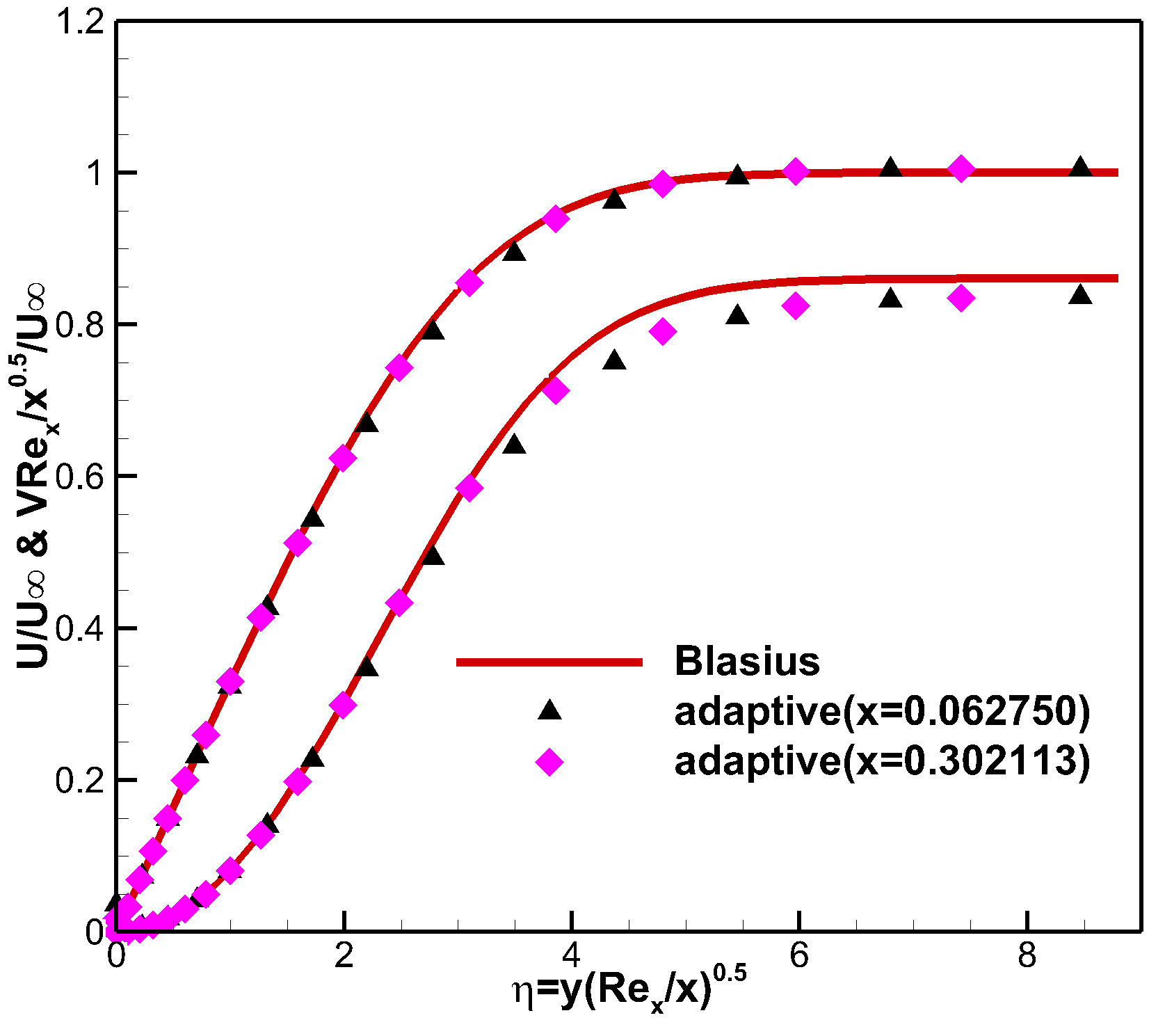}
 \label{plate_uv_adaptive}
 }
	\caption{Comparison between the numerical results and the Blasius solutions for the flat-plate boundary layer}
\end{figure}

First, the hybrid approach without adaptive strategies is employed to perform the flow field analysis, where Eq.~\ref{fhybrid2} is used to evaluate the fluxes at cell interfaces. The velocity space is defined over the range $[-4,4]\times[-4,4]$ and discretized using 17 uniformly distributed velocity points. Numerical quadrature is carried out using the Newton–Cotes method. Figures~\ref{plate_U0} and ~\ref{plate_V0} show the velocity contours in the whole computational domain. To quantitatively compare the numerical results to the Blasius solutions, the velocity profiles along the y direction at two locations of x=0.062750, and x=0.302113 are extracted, as shown in Fig.~\ref{plate_uv_hybrid}. Both U and V have a fairly good agreement with the Blasius solutions.

\begin{figure}[h!]
    \centering
\includegraphics[width=0.7\linewidth]{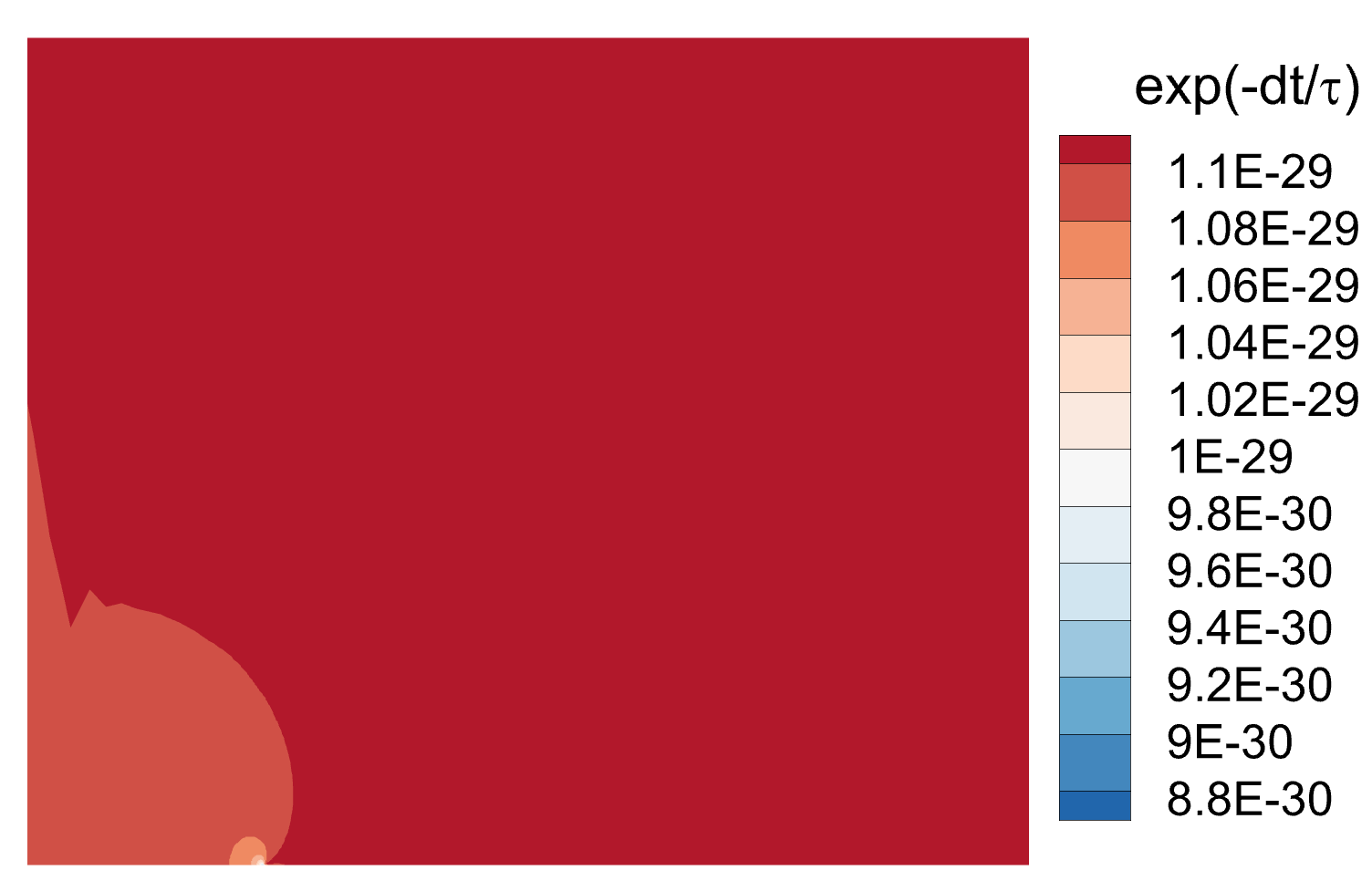}
    \caption{The weighting factors in the whole computational domain}
   \label{plate_dt}
\end{figure}

 In the continuum flow regimes, the time step is much larger than the collision time, i.e., $\Delta t\gg \tau$, and the exponential term $e^{-\Delta t/\tau}$ is therefore expected to be close to zero, as illustrated in Fig.~\ref{plate_dt}. Since $\tau$ is close to $\tau_n$ for smooth continuum flows, the weighting factor $e^{-\Delta t/\tau_n}$ also approaches zero, indicating that the flux contribution from the DVM becomes negligible. To further reduce computational cost, the adaptive strategy introduced in the methodology section is employed in this case. Figure~\ref{plate_uv_adaptive} shows $U$ and $V$ velocity profiles along the vertical direction, where excellent agreement between the adaptive approach and the Blasius solution is observed. In the legend, 'adaptive' denotes the hybrid approach with adaptive strategies.

\begin{figure}[h!]
    \centering
\includegraphics[width=2.25in]{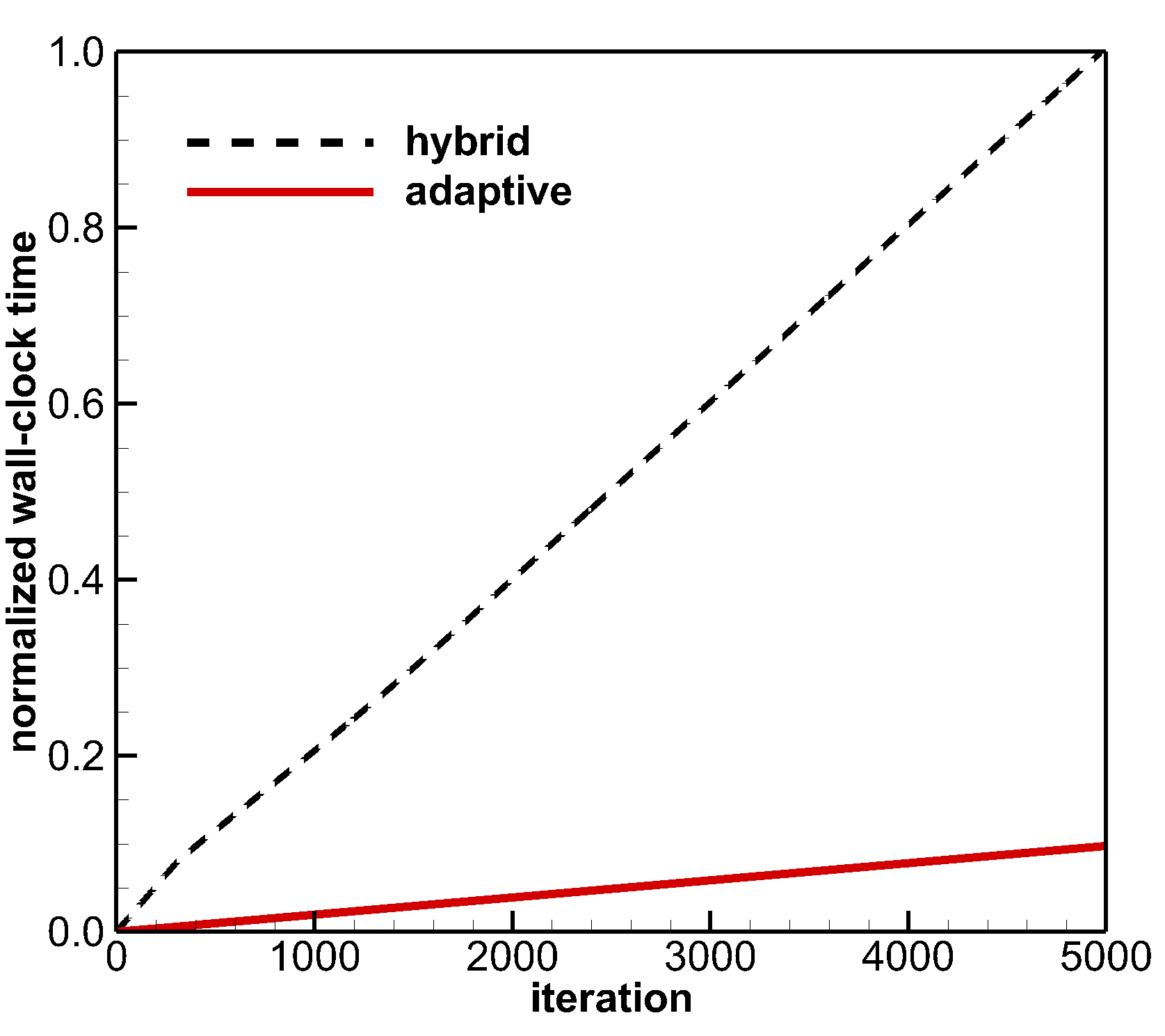}
    \caption{Comparison of wall-clock time between the hybrid approaches with and without adaptive strategies}
   \label{plate_time}
\end{figure}
To demonstrate the computational efficiency of the adaptive strategy over the original hybrid approach, the wall‑clock time is compared. To ensure a fair and reliable comparison, both simulations are performed on the same workstation equipped with an Intel(R) Xeon(R) Gold 6130 CPU $@$ 2.10GHz under identical computational workloads. The comparison of wall‑clock time is presented in Fig.~\ref{plate_time}. Compared with the hybrid approach without adaptivity, the incorporation of the adaptive strategies reduces the computational cost by approximately a factor of ten. It is worth noting that, to improve computational efficiency, the adaptive strategies are adopted in all subsequent case studies.

\subsection{Case 2: Lid-Driven Cavity Flow}

\begin{figure}[h!]
    \centering
\includegraphics[width=0.5\linewidth]{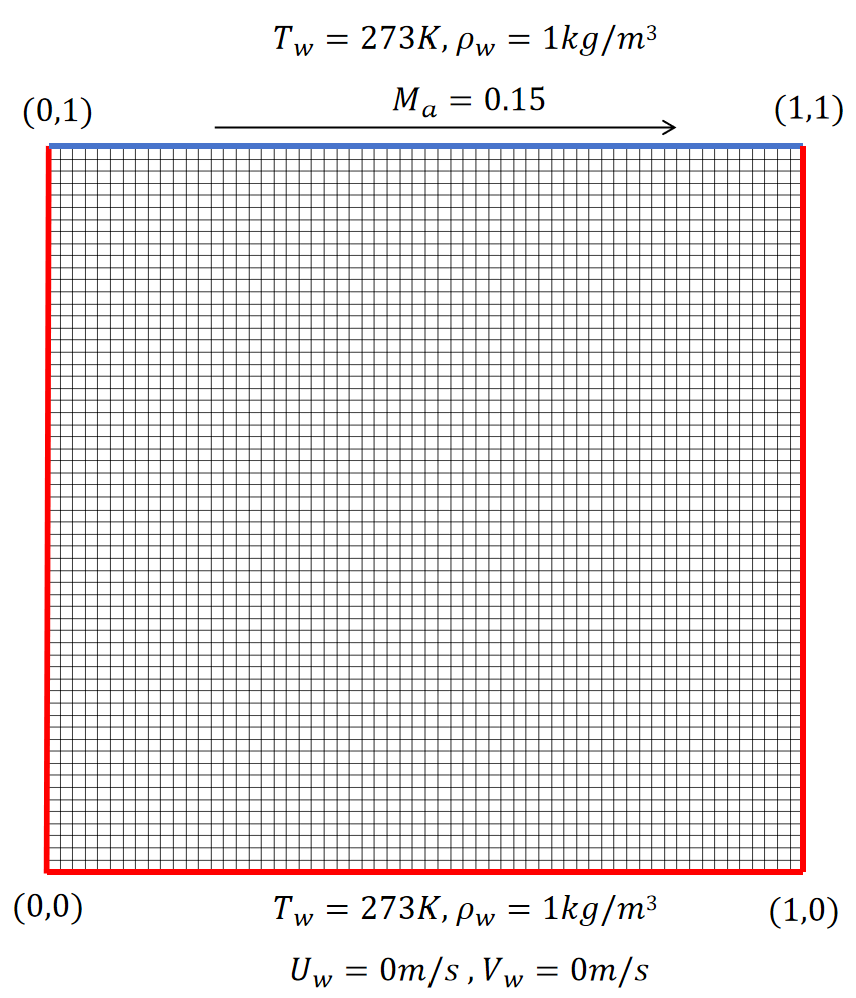}
    \caption{The computational mesh of the lid-driven cavity flow}
   \label{cavity_mesh}
\end{figure}

The second test case is the lid-driven cavity flow. Three different Knudsen numbers are considered: 10.0, 0.075 and $2.05154\times10^{-4}$. Figure~\ref{cavity_mesh} presents the computational mesh for the lid-driven cavity configuration. The computational domain is defined on $[0,1]\times[0,1]$, with a mesh resolution of 61$\times$61.
In the particle velocity space, the Newton-Cotes quadrature with 61$\times$61 velocity points uniformly distributed in $[-4,4]\times[-4,4]$ is employed in the case of $Kn=10$, and the Gauss-Hermite quadrature with 28$\times$28 velocity points is adopted in the case of $Kn=0.075$ and $Kn=2.05154\times10^{-4}$.

The boundary conditions are set as follows:
on the left, right, and bottom boundaries, the isothermal wall boundary condition is imposed, with the wall density and temperature prescribed as $1kg/m^3$ and $273K$, respectively. On the top boundary, the isothermal slip wall boundary condition is employed, while the wall temperature is also maintained at $273K$. The flow goes from left to right, corresponding to a Mach number of 0.15. The Prandtl number is set to 2/3 and the working flow is Argon. For the free-molecular and slip cases, the reference viscosity is determined by Eq.~\ref{Kn}, while for the continuous case, the reference viscosity is calculated by
\begin{equation}
\mu_{ref} = \frac{\rho_w U_w L}{Re}
\end{equation}

\begin{figure}[h!]
\centering
\subfigure[U]{
	\includegraphics[width=4.0in]{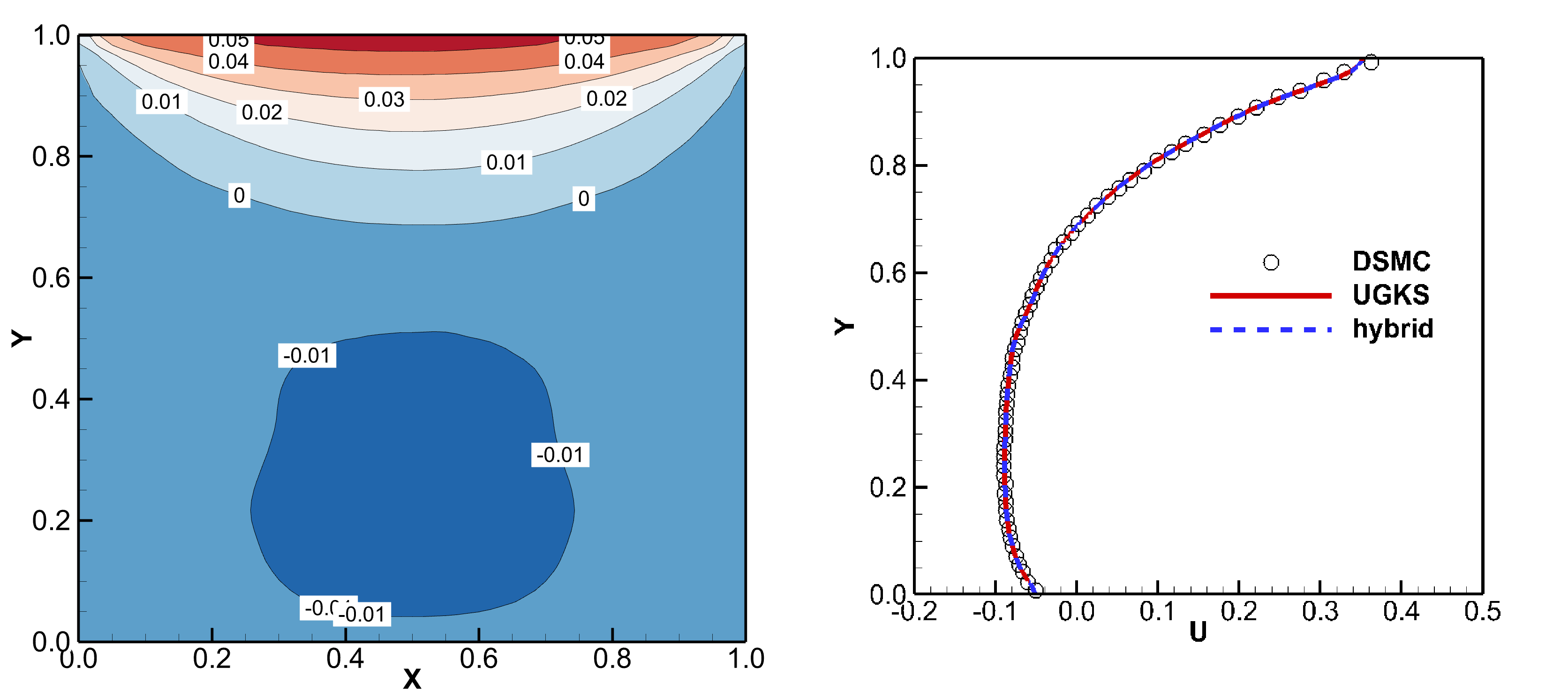}
 \label{cavity_Kn10_U}
 }
 \subfigure[V]{
	\includegraphics[width=4.0in]{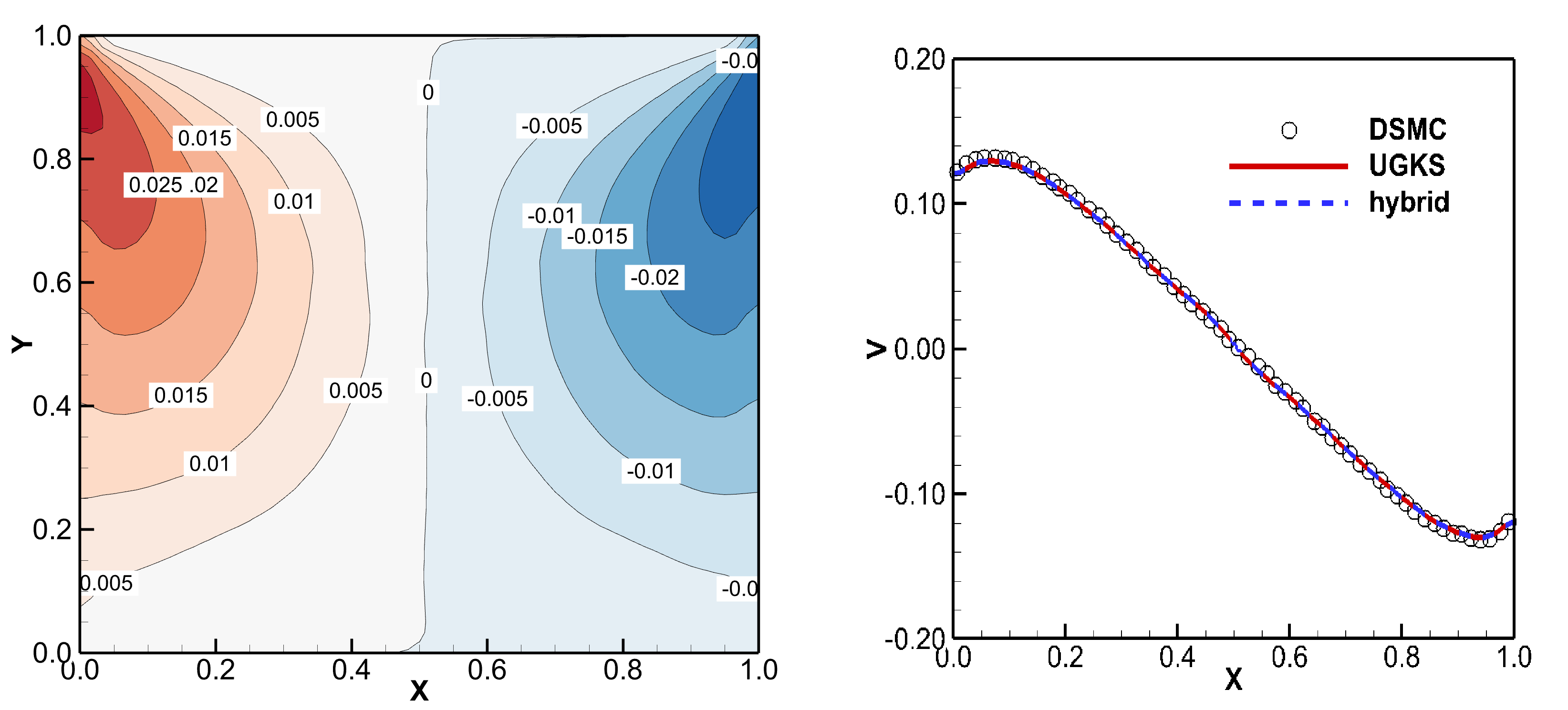}
 \label{cavity_Kn10_V}
 }
	\caption{The velocity contours in the whole computational domain (UGKS: the colored background; hybrid: the solid lines) and velocity along the central line at $Kn=10$: a) U; b) V}
\end{figure}

\begin{figure}[h!]
\centering
\subfigure[residual]{
	\includegraphics[width=2.0in]{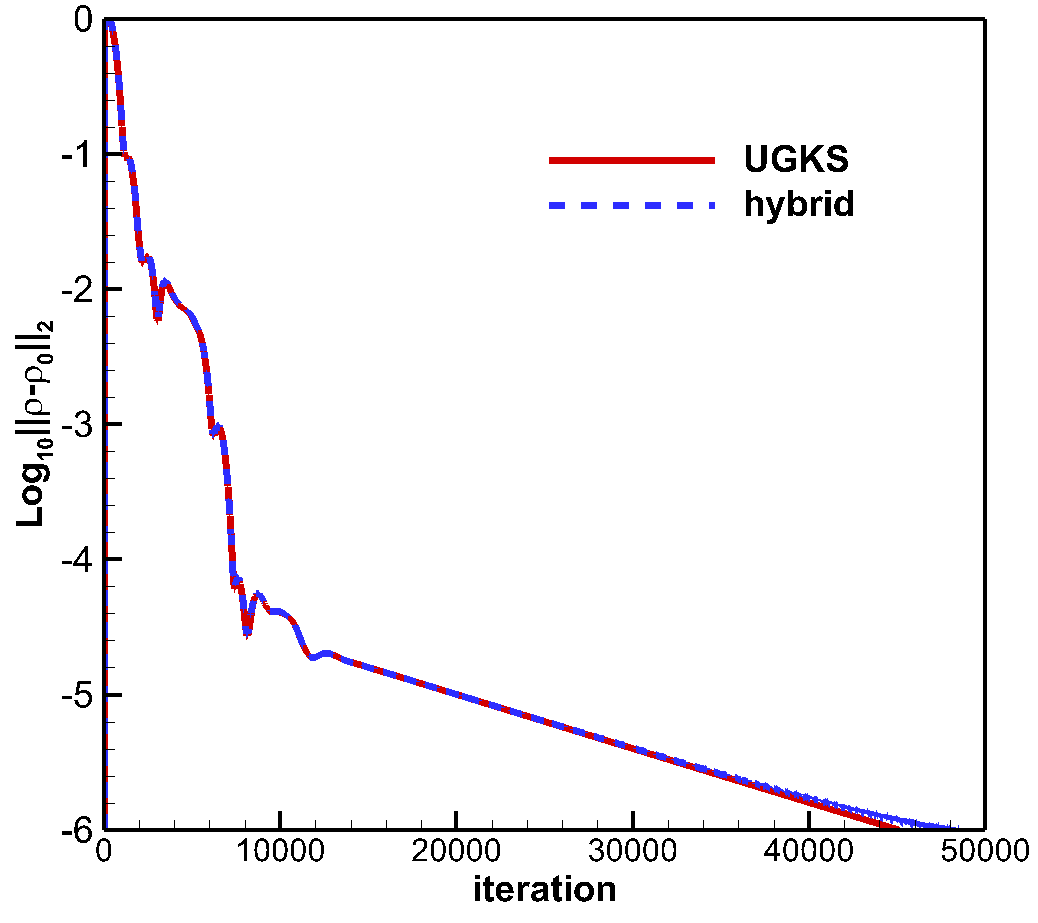}
 \label{cavity_Kn10_res}
 }
 \subfigure[wall-clock time]{
	\includegraphics[width=2.0in]{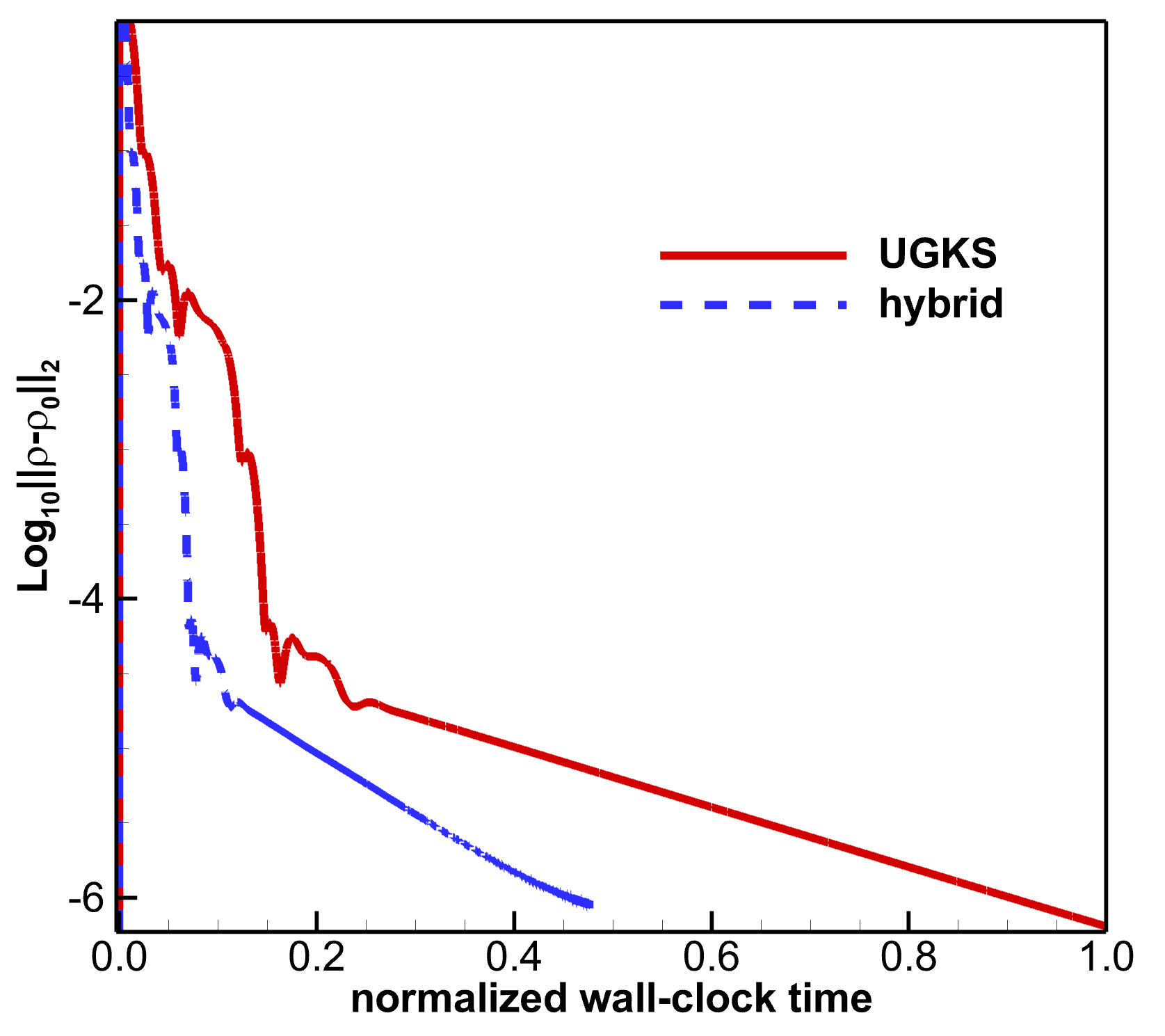}
 \label{cavity_Kn10_time}
 }
	\caption{The density residual at $Kn=10$: a) convergence histories; b) wall-clock time}
\end{figure}

Figures~\ref{cavity_Kn10_U} and~\ref{cavity_Kn10_V} present comparisons of the $U$- and $V$-velocity contours and the corresponding centerline profiles at $Kn=10$. In the legends, “DSMC” denotes results obtained using the Direct Simulation Monte Carlo (DSMC) method, which serves as the reference solution~\cite{JOHN2011197}. In the contour plots, the colored backgrounds correspond to the UGKS solutions, while the solid lines represent the results from the hybrid approach. Overall, excellent agreement is observed among the UGKS, hybrid solutions, and the DSMC benchmarks.
Figure~\ref{cavity_Kn10_res} compares the convergence histories of the density residuals, while Fig.~\ref{cavity_Kn10_time} presents a comparison of the wall-clock time between the UGKS and the hybrid approach. Both methods exhibit nearly identical convergence behaviors; however, the hybrid approach requires only about half of the wall-clock time of the UGKS, demonstrating its significantly improved computational efficiency for rarefied flow simulations.

\begin{figure}[h!]
\centering
\subfigure[U]{
	\includegraphics[width=3.6in]{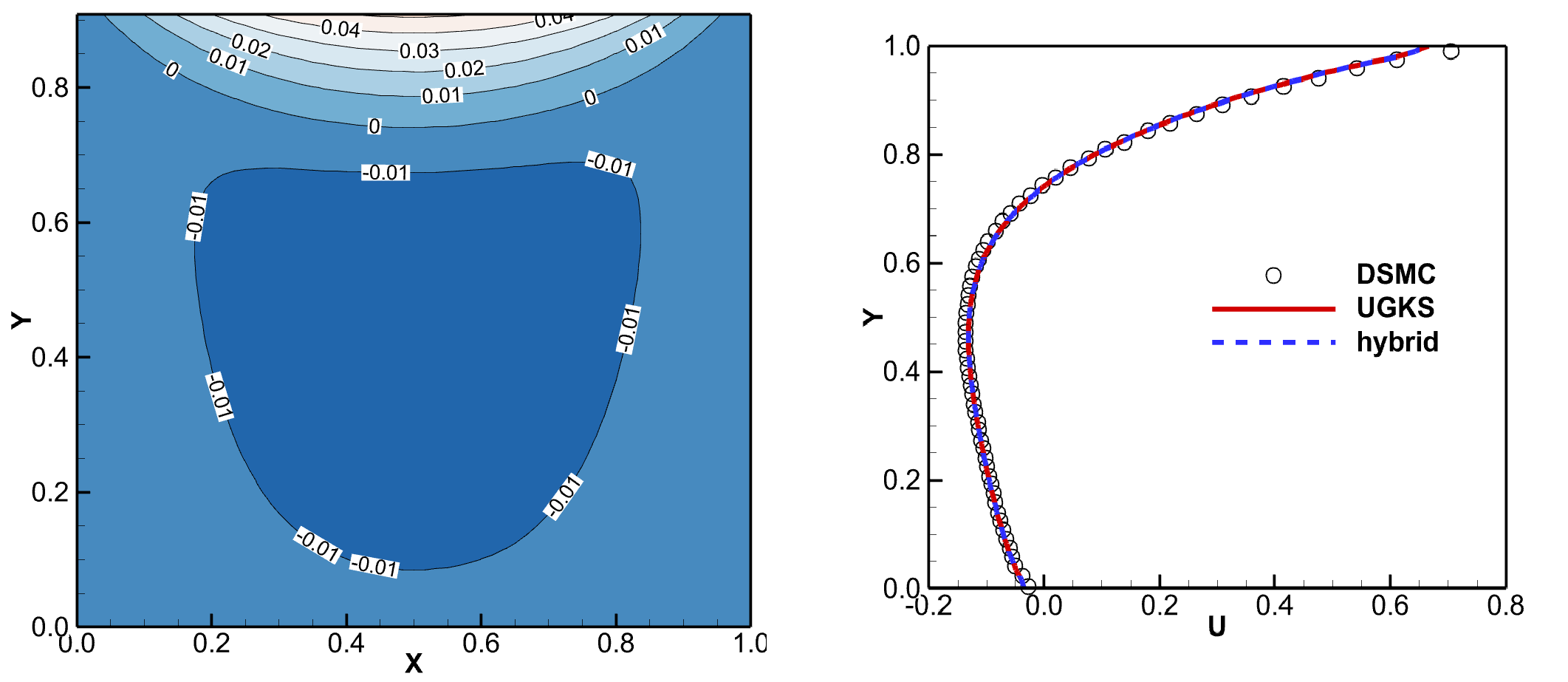}
 \label{cavity_Kn0.075_U}
 }
 \subfigure[V]{
	\includegraphics[width=3.6in]{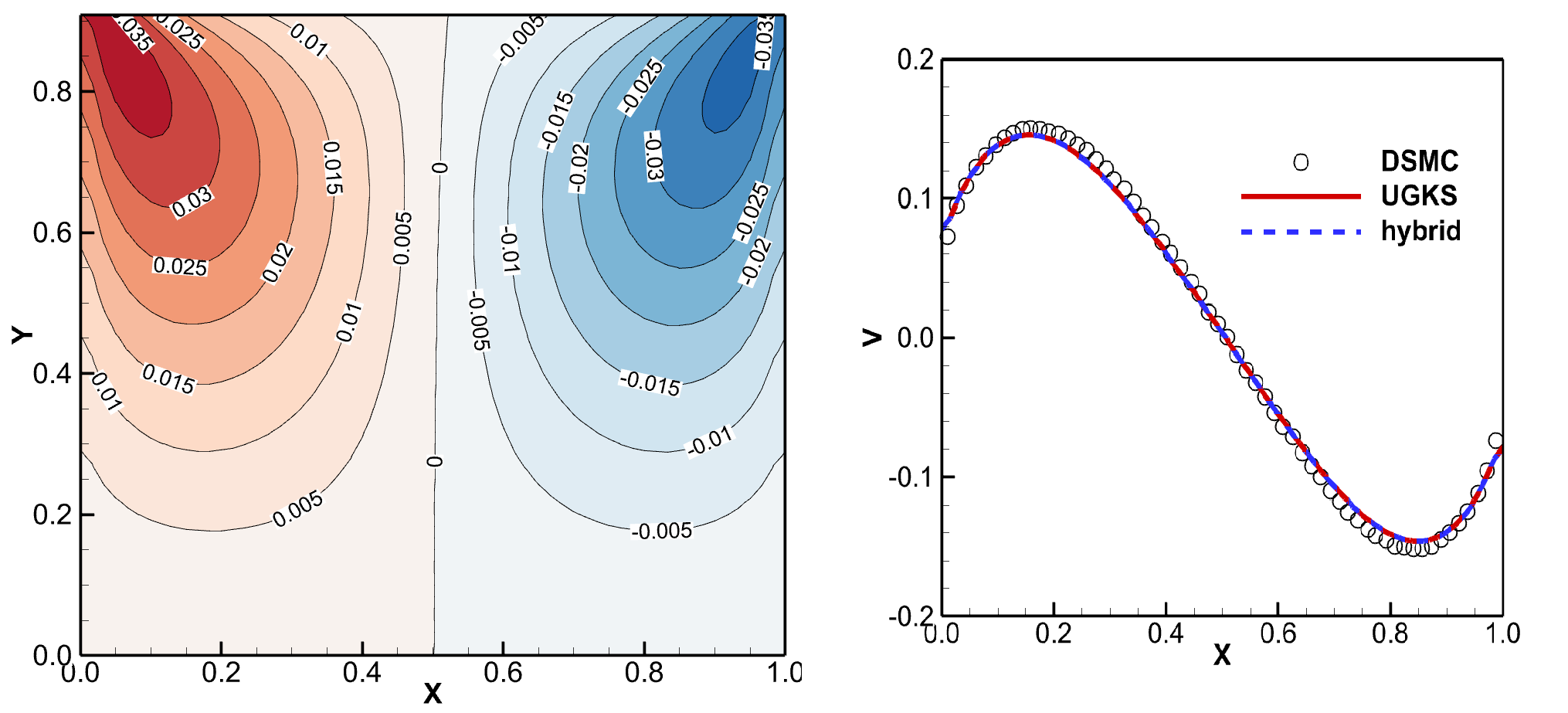}
 \label{cavity_Kn0.075_V}
 }
	\caption{The velocity contours in the whole computational domain (UGKS: the colored background; hybrid: the solid lines) and velocity along the central line at $Kn=0.075$: a) U; b) V}
\end{figure}

\begin{figure}[h!]
\centering
\subfigure[residual]{
	\includegraphics[width=2.0in]{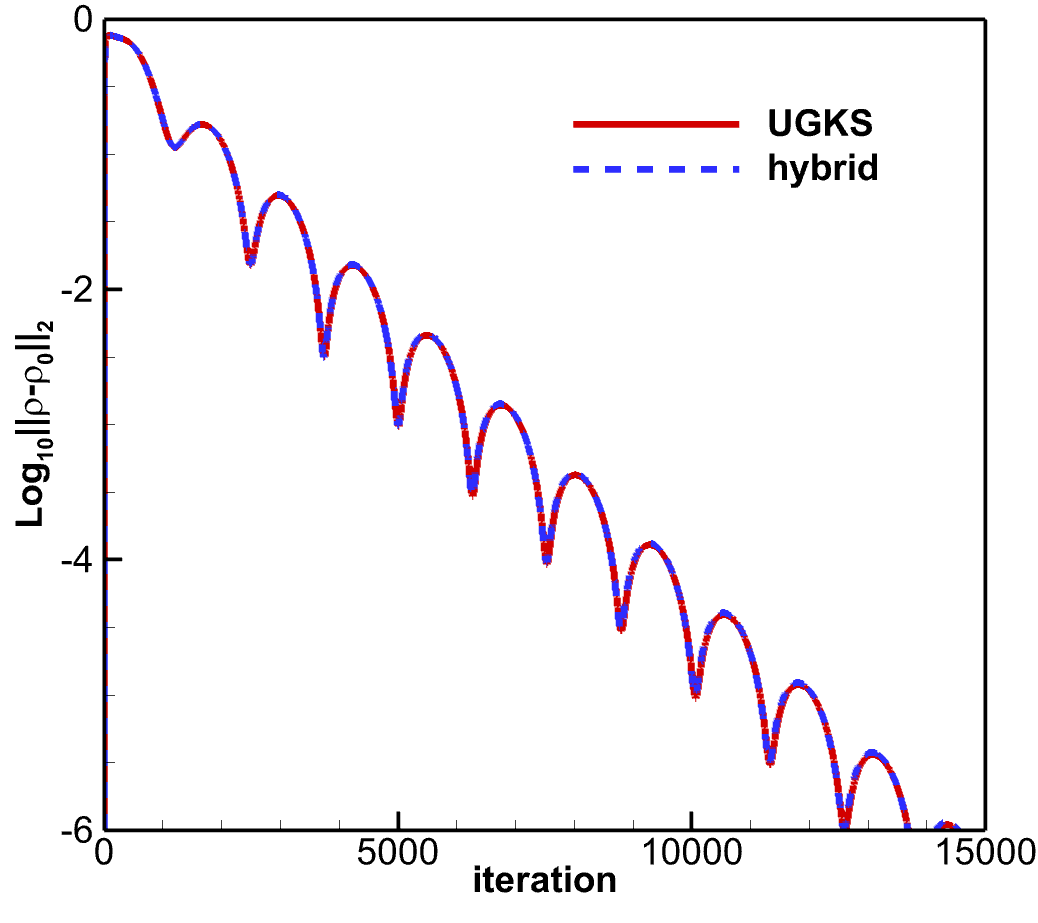}
 \label{cavity_Kn0.075_res}
 }
 \subfigure[wall-clock time]{
	\includegraphics[width=2.0in]{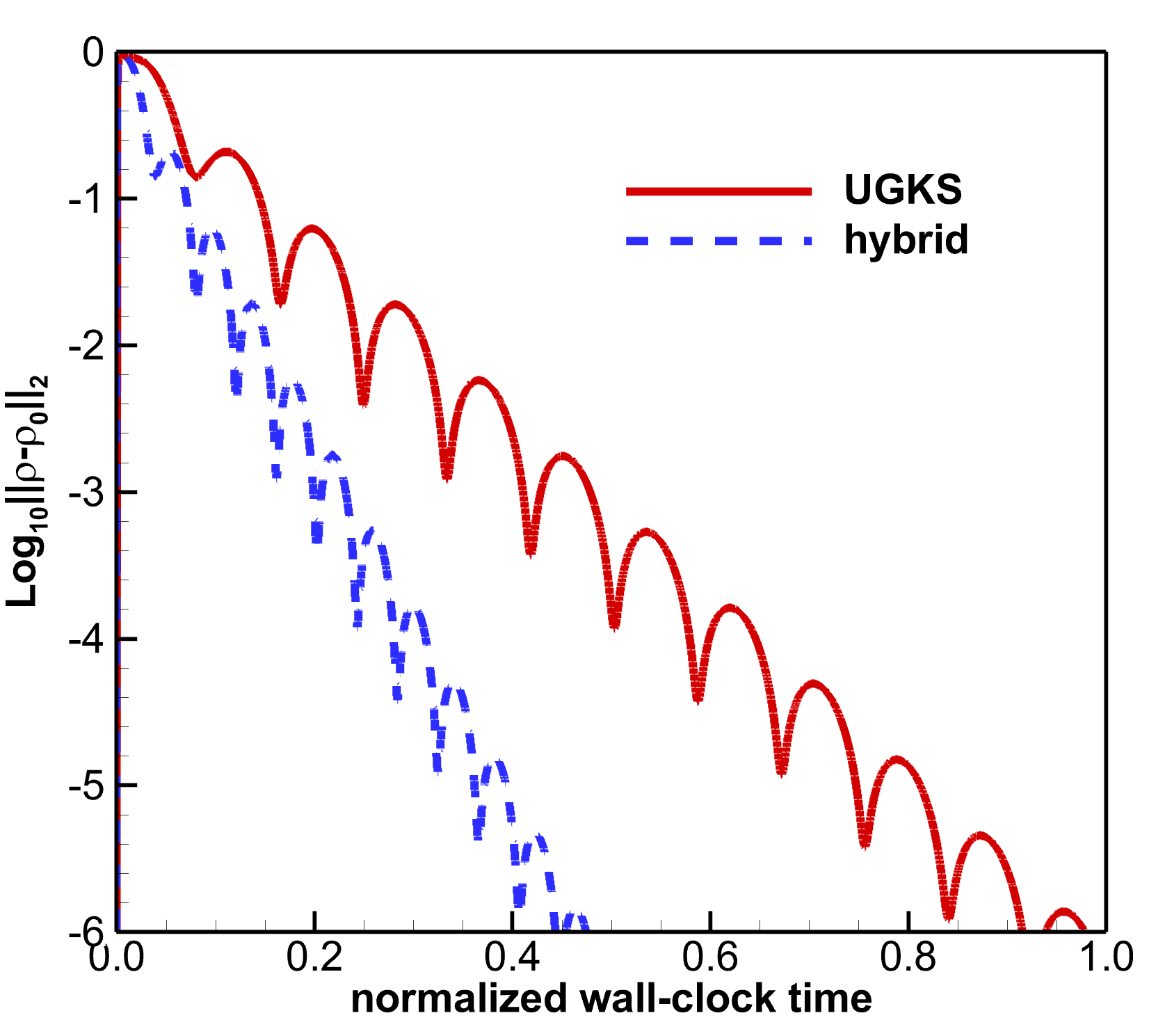}
 \label{cavity_Kn0.075_time}
 }
	\caption{The density residual at $Kn=0.075$: a) convergence histories; b) wall-clock time}
\end{figure}
To further assess the effectiveness of the hybrid approach in the slip flow regime, the Knudsen number is reduced from 10 to 0.075. Similar to the case at $Kn=10$, the hybrid approach achieves solution accuracies comparable to those of the UGKS and the DSMC, as shown in Figs.~\ref{cavity_Kn0.075_U} and ~\ref{cavity_Kn0.075_V}, while exhibiting a significantly reduced computational cost. In particular, the wall-clock time is reduced by approximately a factor of 2.3 compared to the UGKS (see Figs.~\ref{cavity_Kn0.075_time} and ~\ref{cavity_Kn0.075_time}), demonstrating the superior computational efficiency of the hybrid approach at moderate rarefaction levels.

\begin{figure}[h!]
\centering
\subfigure[U]{
	\includegraphics[width=2.25in]{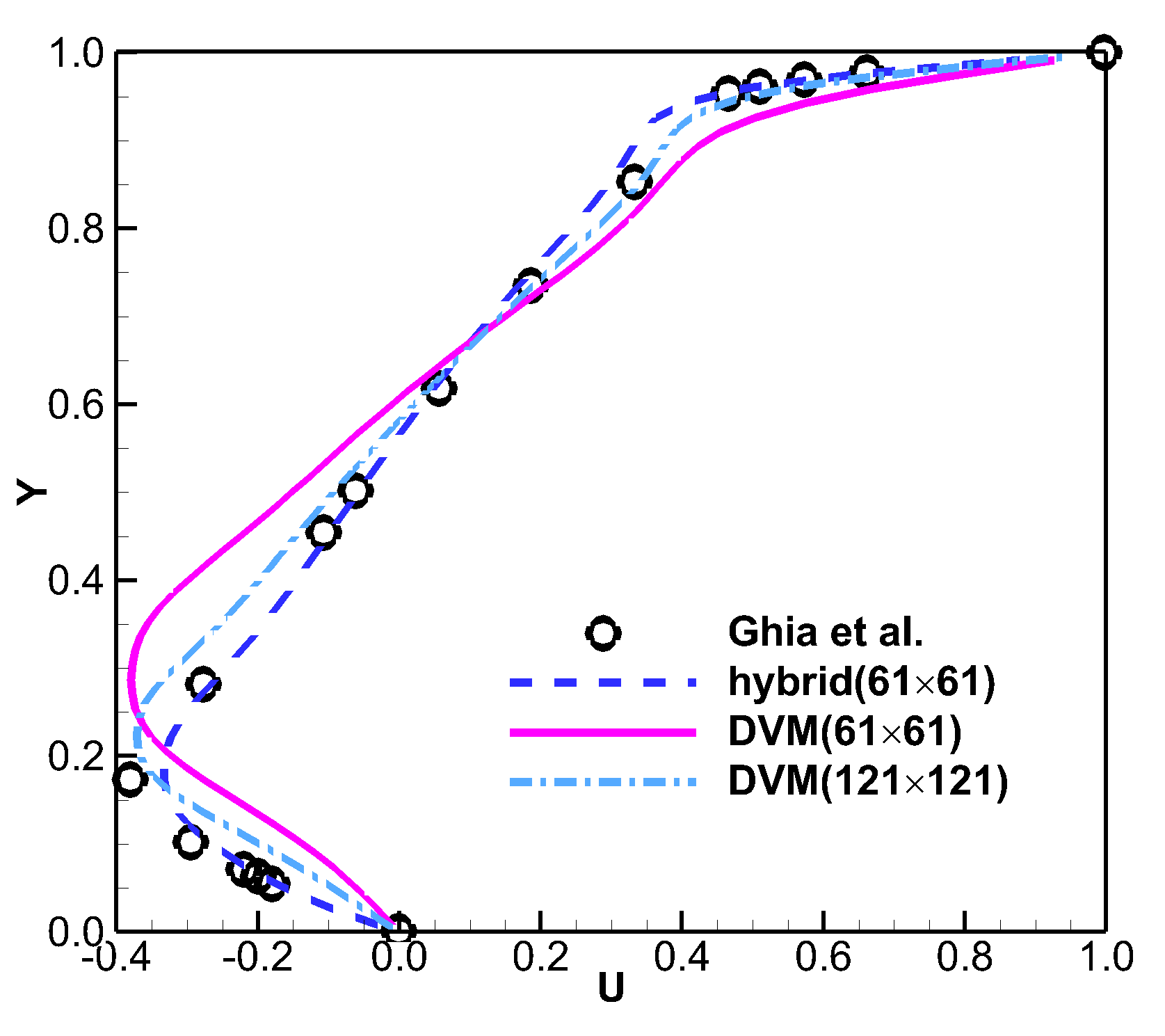}
 \label{cavity_Re1000_Uy}
 }
 \subfigure[V]{
	\includegraphics[width=2.25in]{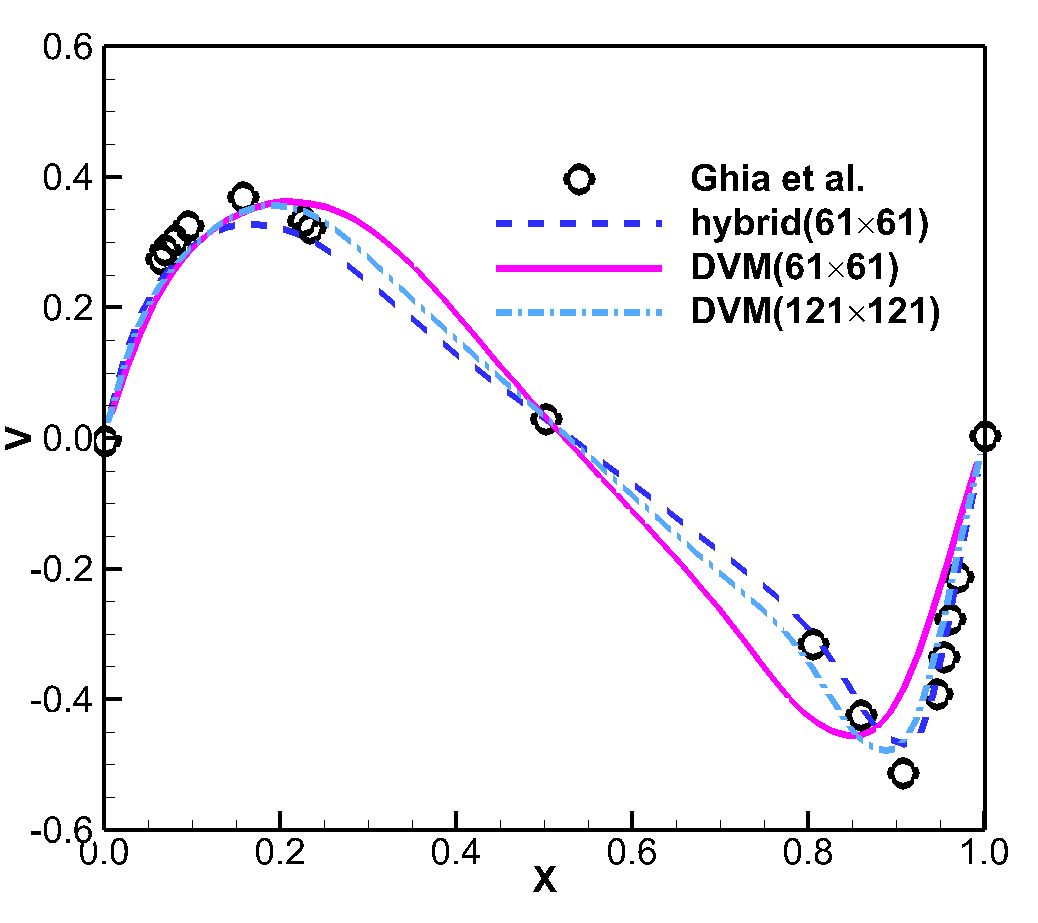}
 \label{cavity_Re1000_Vx}
 }
	\caption{The velocity profiles along the central line at $Kn=2.05154\times10^{-4}$: a) horizontal; b) vertical}
\end{figure}
\begin{figure}[h!]
\centering
\subfigure[DVM]{
	\includegraphics[width=2.25in]{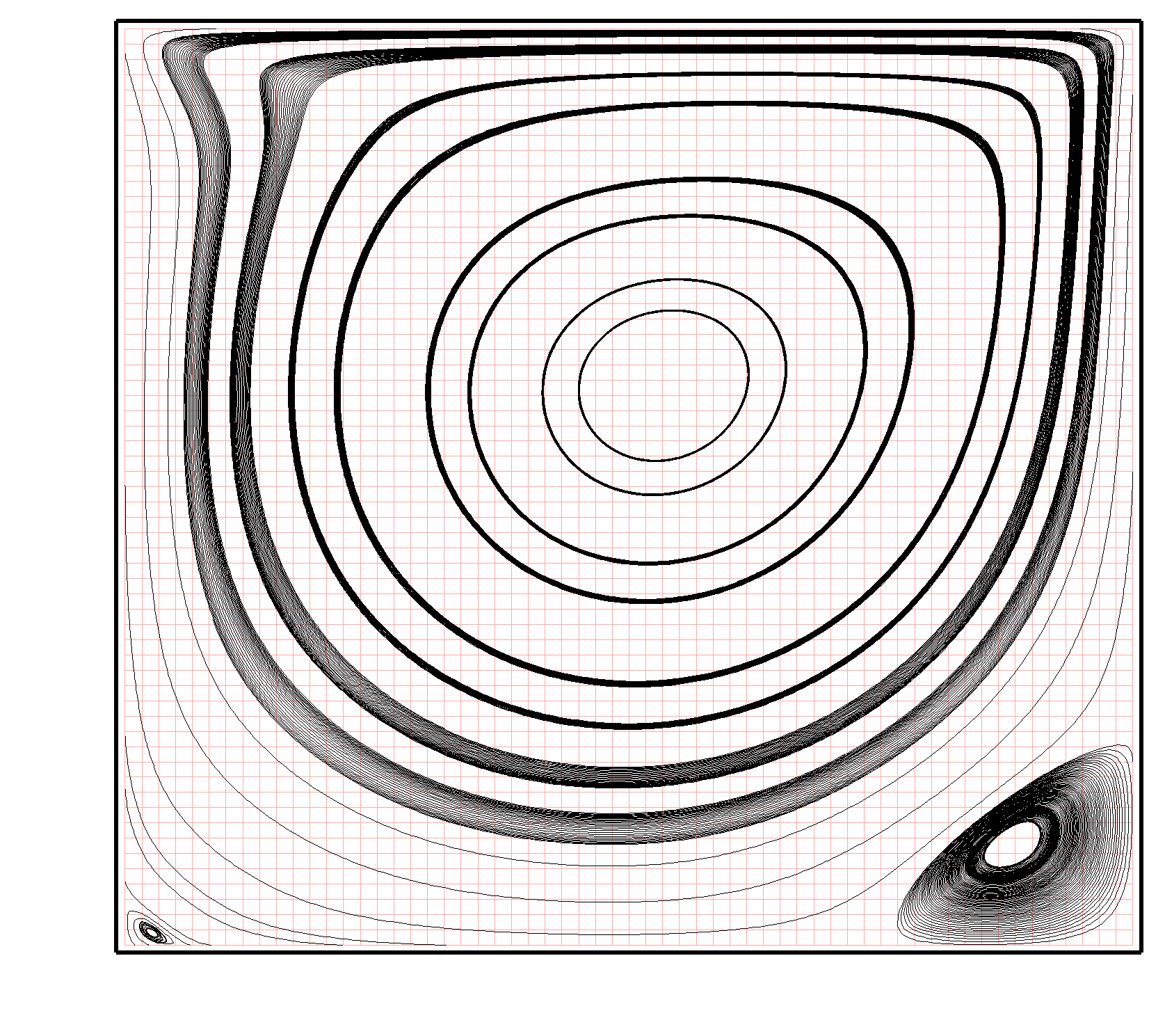}
 \label{cavity_Re1000_stream_dvm61}
 }
 \subfigure[hybrid]{
	\includegraphics[width=2.25in]{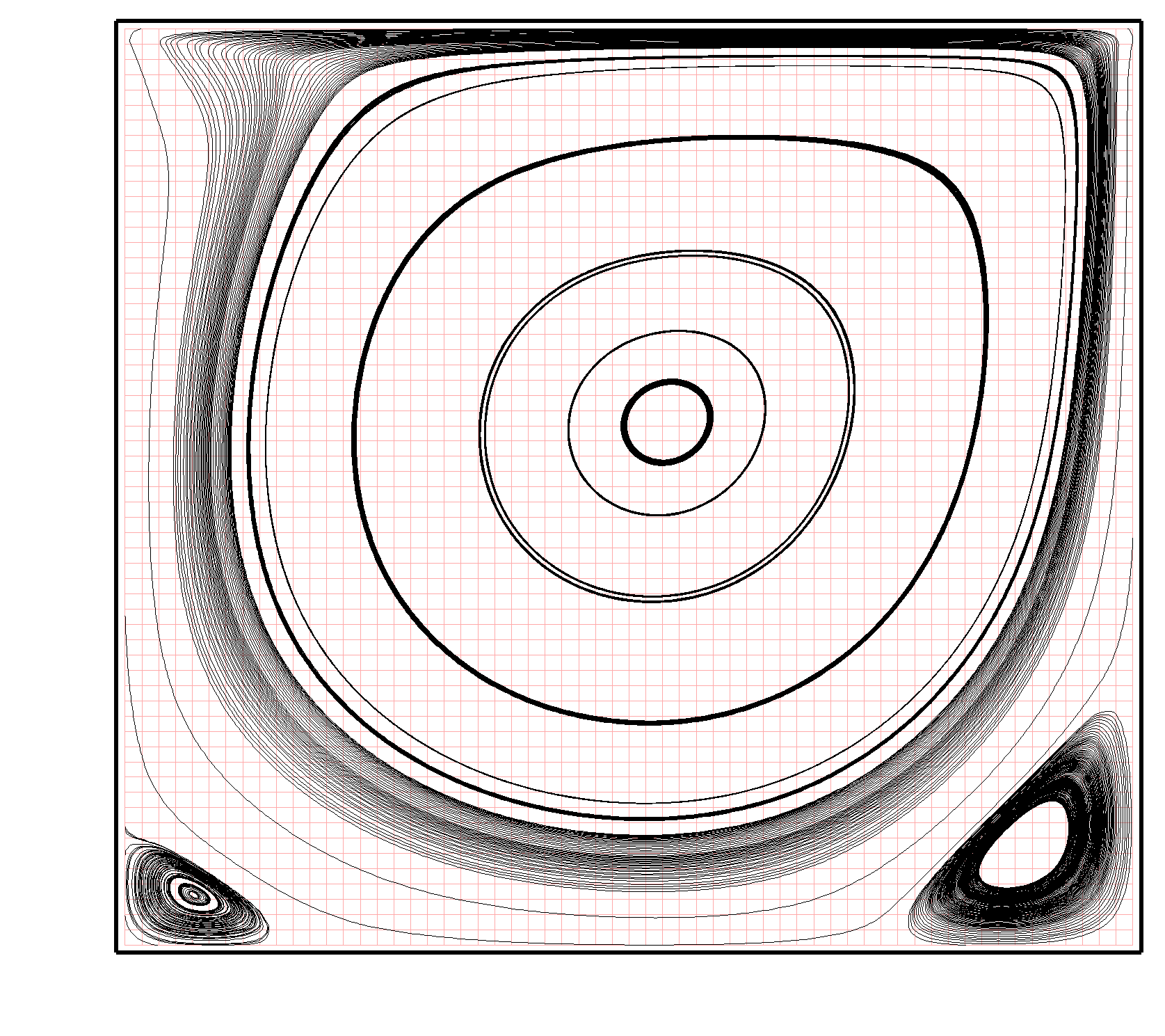}
 \label{cavity_Re1000_stream_hybrid}
 }
	\caption{The streamlines with a mesh resolution of 61$\times$61 at $Kn=2.05154\times10^{-4}$: a) DVM ; b) hybrid}
\end{figure}

When the Knudsen number is further reduced to $Kn=2.05154\times10^{-4}$, corresponding to a Reynolds number of 1000, the flow enters the continuum regime. Figures~\ref{cavity_Re1000_Uy} and~\ref{cavity_Re1000_Vx} present the profiles of the velocity components $U$ and $V$ along the centerlines. The benchmark solutions reported by Ghia et al.~\cite{GHIA1982387} are adopted as reference data.
It can be observed that, with a mesh resolution of 61$\times$61, the results obtained using the hybrid approach agree well with the reference solutions, whereas the DVM results exhibit deviations. When the mesh resolution is increased to 121$\times$121, the accuracy of the DVM is substantially improved, yielding results much closer to the reference data.
Figures~\ref{cavity_Re1000_stream_dvm61} and~\ref{cavity_Re1000_stream_hybrid} compare the corresponding streamline patterns. Notably, in the bottom-left corner of the cavity, the DVM predicts a weaker vortex structure than the hybrid method. This discrepancy can be attributed to the excessive numerical dissipation introduced by the upwind reconstruction for flux evaluation.

\subsection{Case 3: Shock Structures}
The third test case considers one-dimensional shock structures to assess the shock-capturing capability of the proposed hybrid approach. The working fluid is Argon. The pre-shock Mach number is 2.0 and the Prandtl number is set to 2/3. The computational domain has a length of 50 and is uniformly discretized with 400 grid points. The velocity space spans from -15 to 15 and is discretized using 101 velocity points. Three Knudsen numbers, namely 1.0, 0.1, and 0.001, are investigated.
In general, different velocity spaces should be employed for different Knudsen numbers to improve computational efficiency and ensure solution accuracy. However, for simplicity, the same velocity space is adopted for all Knudsen numbers in the present study. The corresponding solutions obtained by the UGKS are used as reference results.

\begin{figure}[h!]
\centering
\subfigure[$\rho$]{
	\includegraphics[width=2.2in]{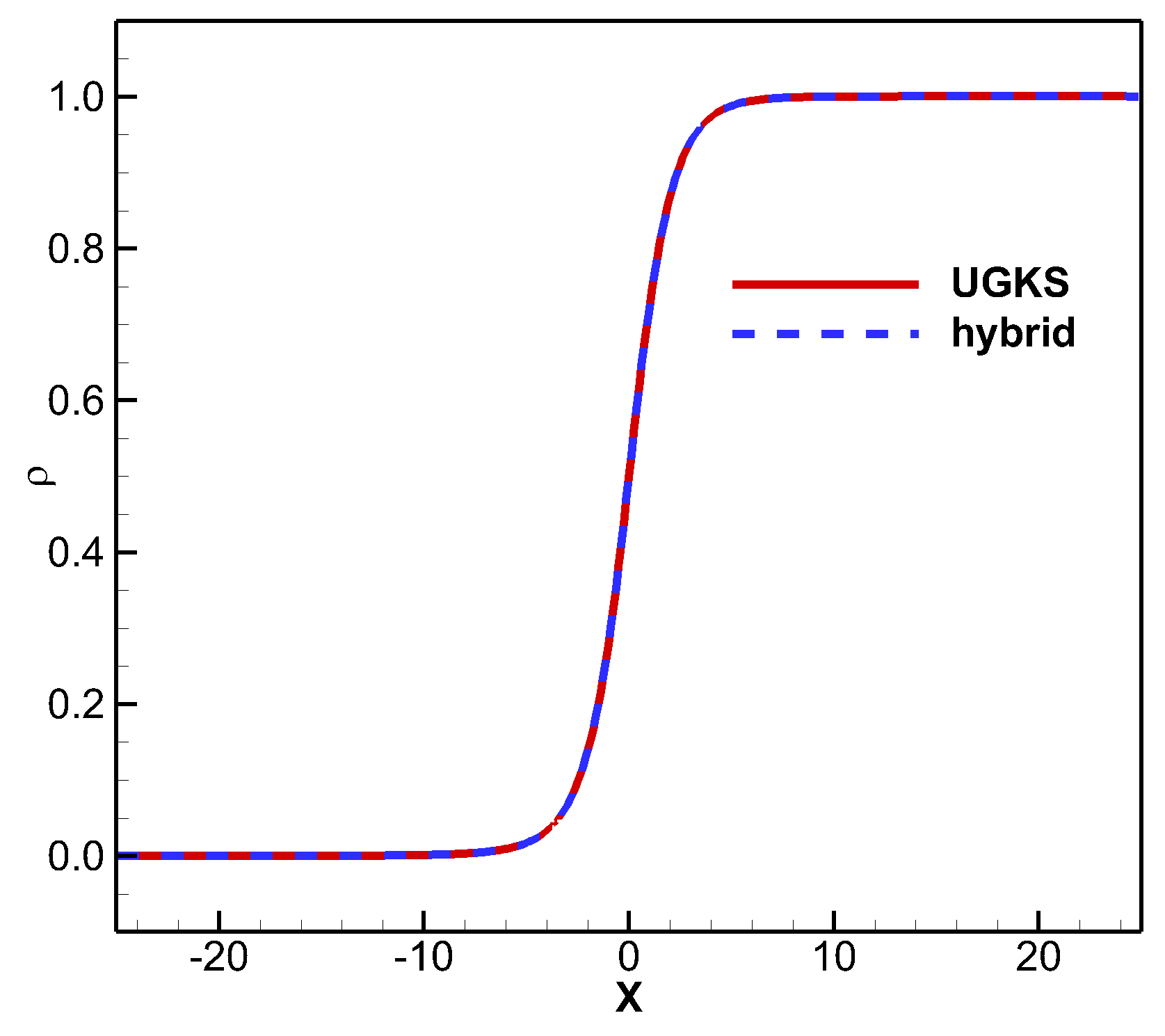}
 \label{shock_Kn1_rho}
 }
 \subfigure[T]{
	\includegraphics[width=2.2in]{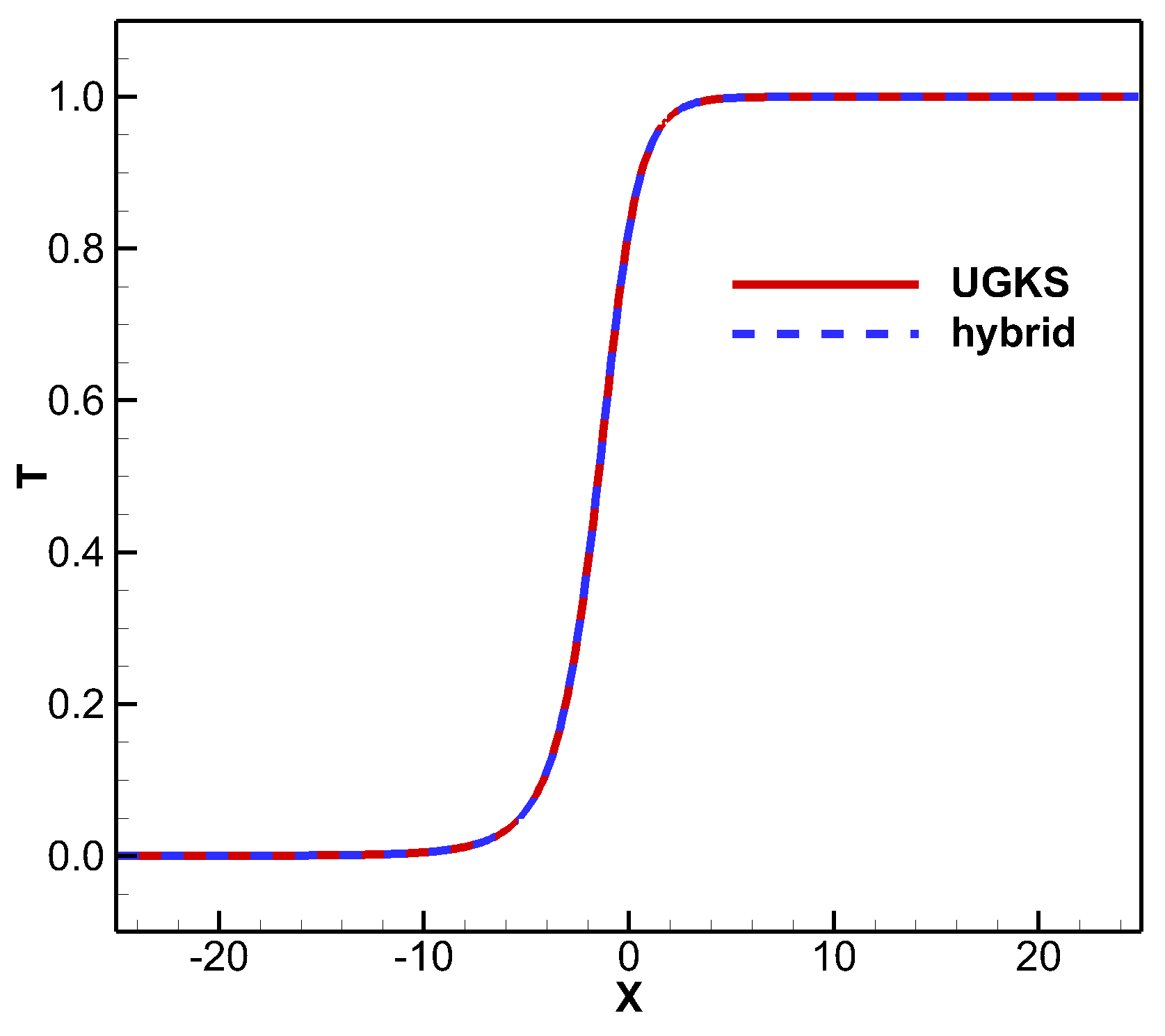}
 \label{shock_Kn1_T}
 }
	\caption{The axial distribution of density and temperature for the shock-structure case at $Kn=1.0$}
\end{figure}
\begin{figure}[h!]
\centering
\subfigure[weighting factor]{
	\includegraphics[width=2.2in]{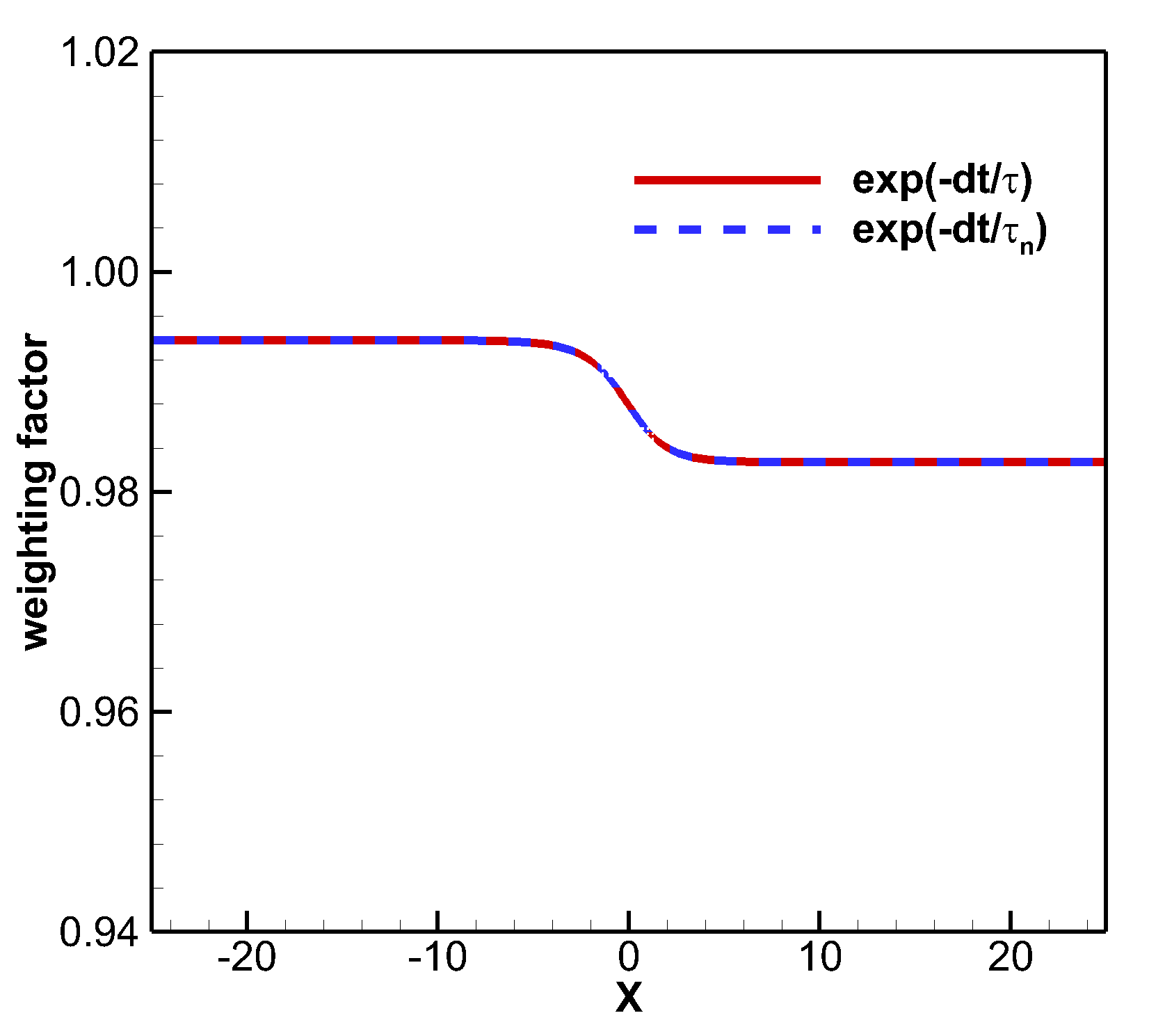}
 \label{shock_Kn1_factor}
 }
 \subfigure[wall-clock time]{
	\includegraphics[width=2.2in]{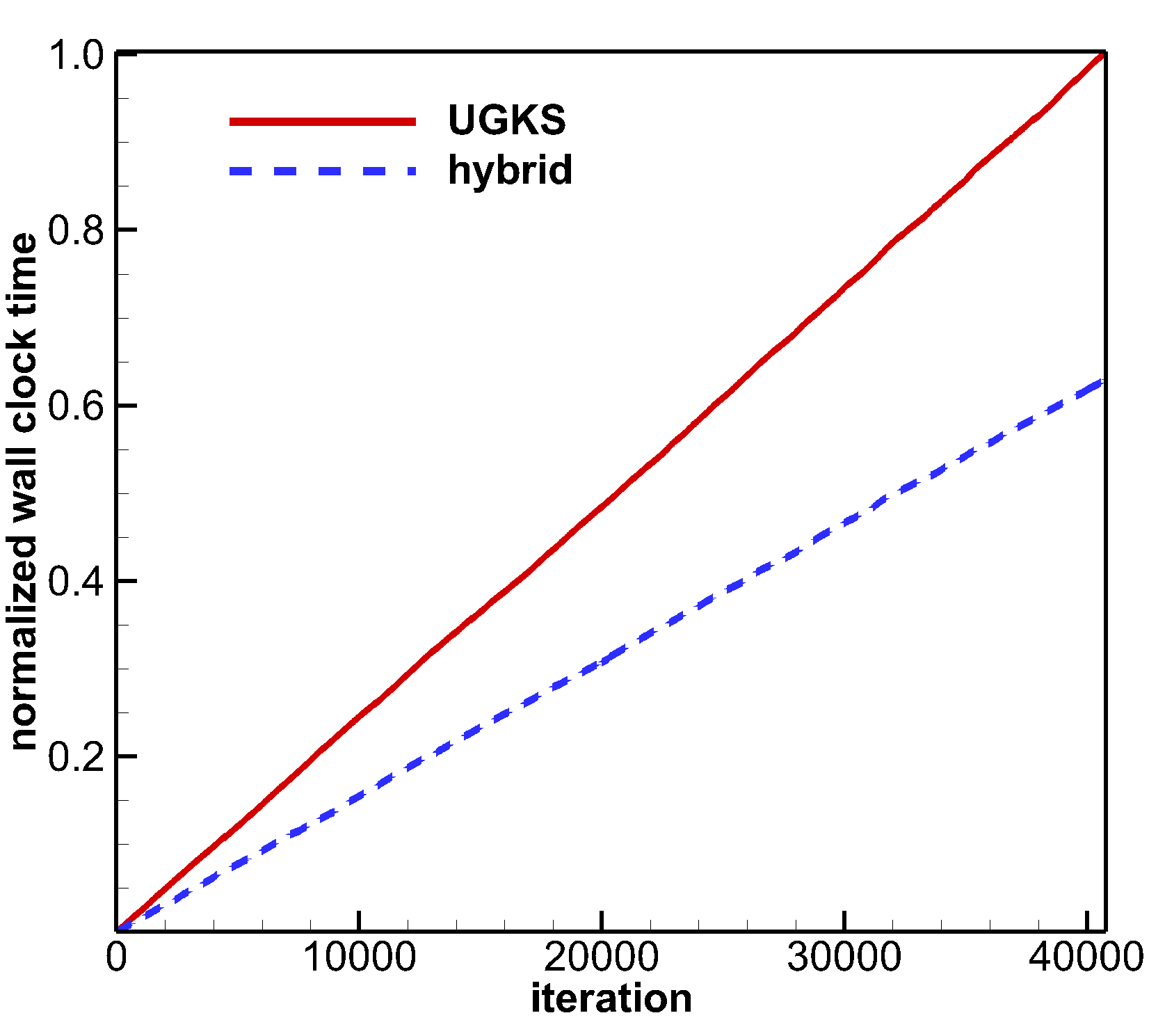}
 \label{shock_Kn1_time}
 }
	\caption{The axial distribution of the weighting factors and comparison of computational cost for the shock-structure case at $Kn=1.0$}
\end{figure}

Figures~\ref{shock_Kn1_rho} and~\ref{shock_Kn1_T} show the axial distributions of density and temperature for the shock-structure case at ${Kn}=1.0$. The results demonstrate that the proposed hybrid approach can accurately resolve the shock structure and exhibits good agreement with the UGKS solutions. Figure~\ref{shock_Kn1_factor} presents the axial distribution of the weighting factor. It is observed that over most of the computational domain, the weighting factor $\exp(-\Delta t/\tau_n)$, based on the numerical collision time, remains close to unity, indicating that the DVM flux is dominant. This behavior confirms the validity of the adaptive strategy introduced in Eq.~\ref{hybrid_f}. Moreover, in the transition flow regime, the numerical collision time coincides with the physical collision time. Figure~\ref{shock_Kn1_time} compares the computational cost of the hybrid approach with that of the UGKS, showing that the hybrid approach reduces the computational time by nearly a factor of 1.6.

\begin{figure}[h!]
\centering
\subfigure[$\rho$]{
	\includegraphics[width=2.2in]{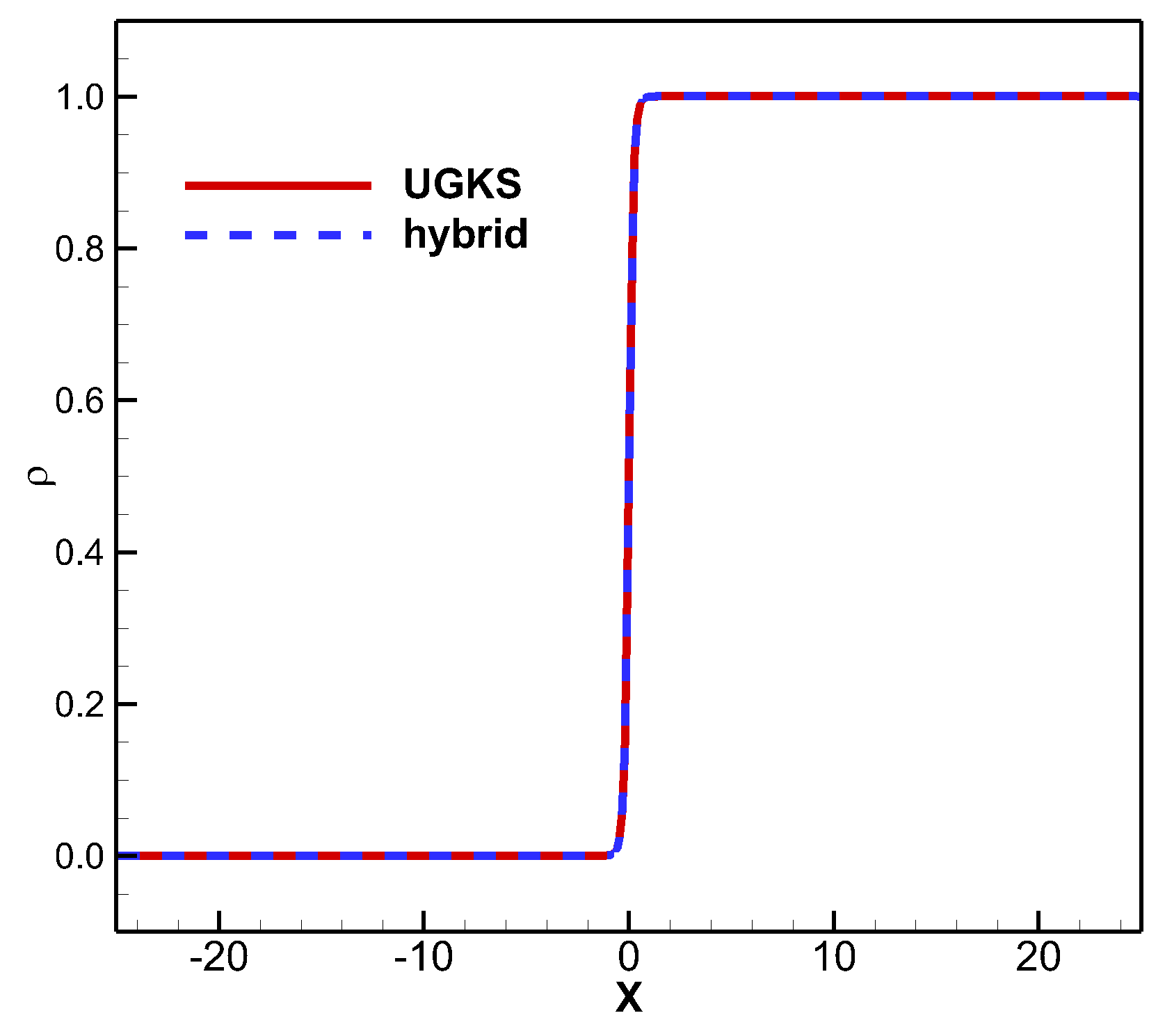}
 \label{shock_Kn0.1_rho}
 }
 \subfigure[T]{
	\includegraphics[width=2.2in]{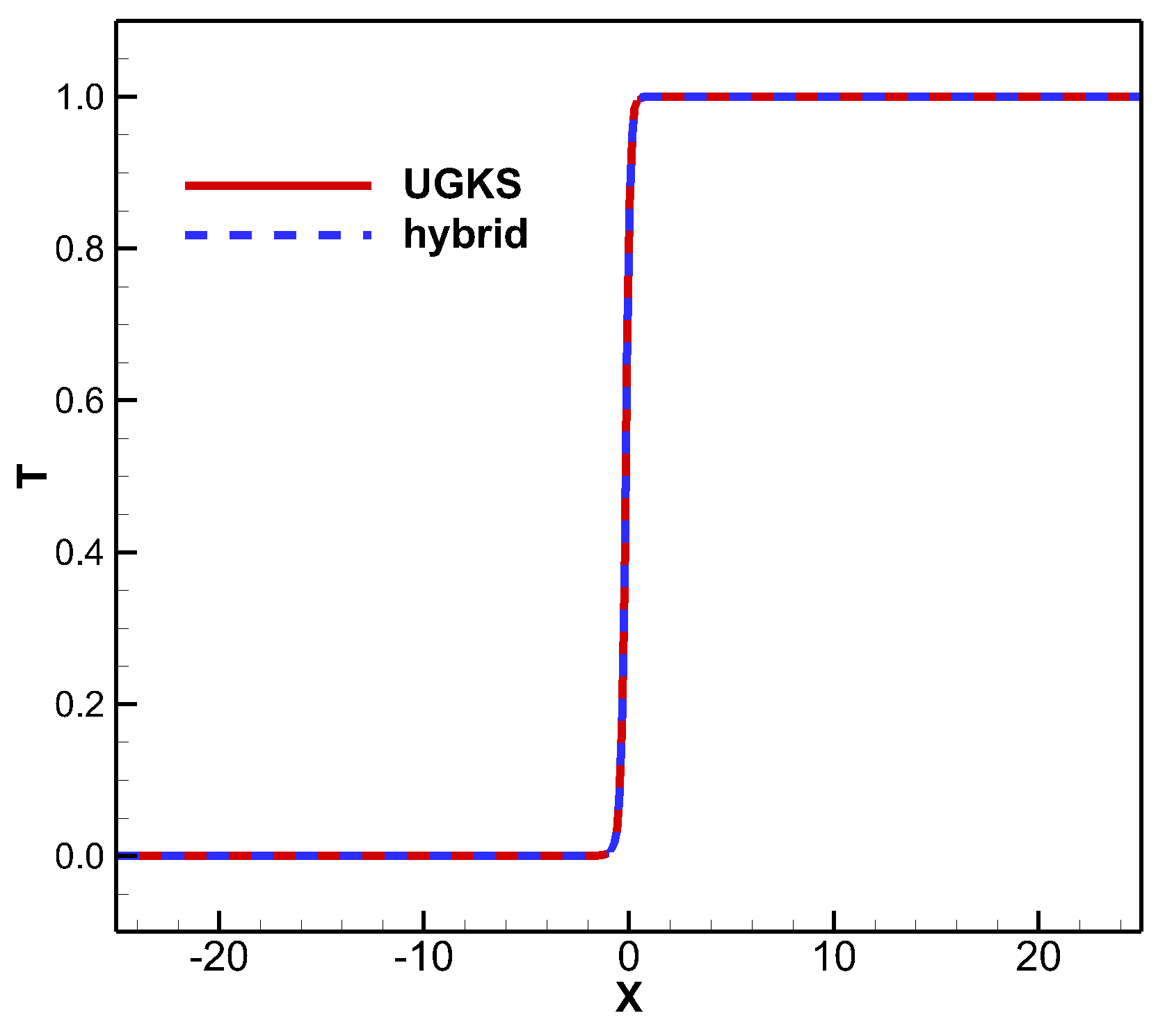}
 \label{shock_Kn0.1_T}
 }
	\caption{The axial distribution of density and temperature for the shock-structure case at $Kn=0.1$}
\end{figure}

The Knudsen number is decreased from 1.0 to 0.1. Similar to the case at ${Kn}=1.0$, excellent agreement in the flow variables between the hybrid approach and the UGKS is observed, as shown in Figs.~\ref{shock_Kn0.1_rho} and~\ref{shock_Kn0.1_T}. With the decrease in the Knudsen number, the weighting factor slightly decreases; however, it remains larger than 0.84 throughout the computational domain. This indicates that the DVM flux is still dominant, further demonstrating the effectiveness of the proposed adaptive strategy. In addition, the computational cost of the UGKS is approximately 1.7 times that of the hybrid approach.
\begin{figure}[h!]
\centering
\subfigure[weighting factor]{
	\includegraphics[width=2.2in]{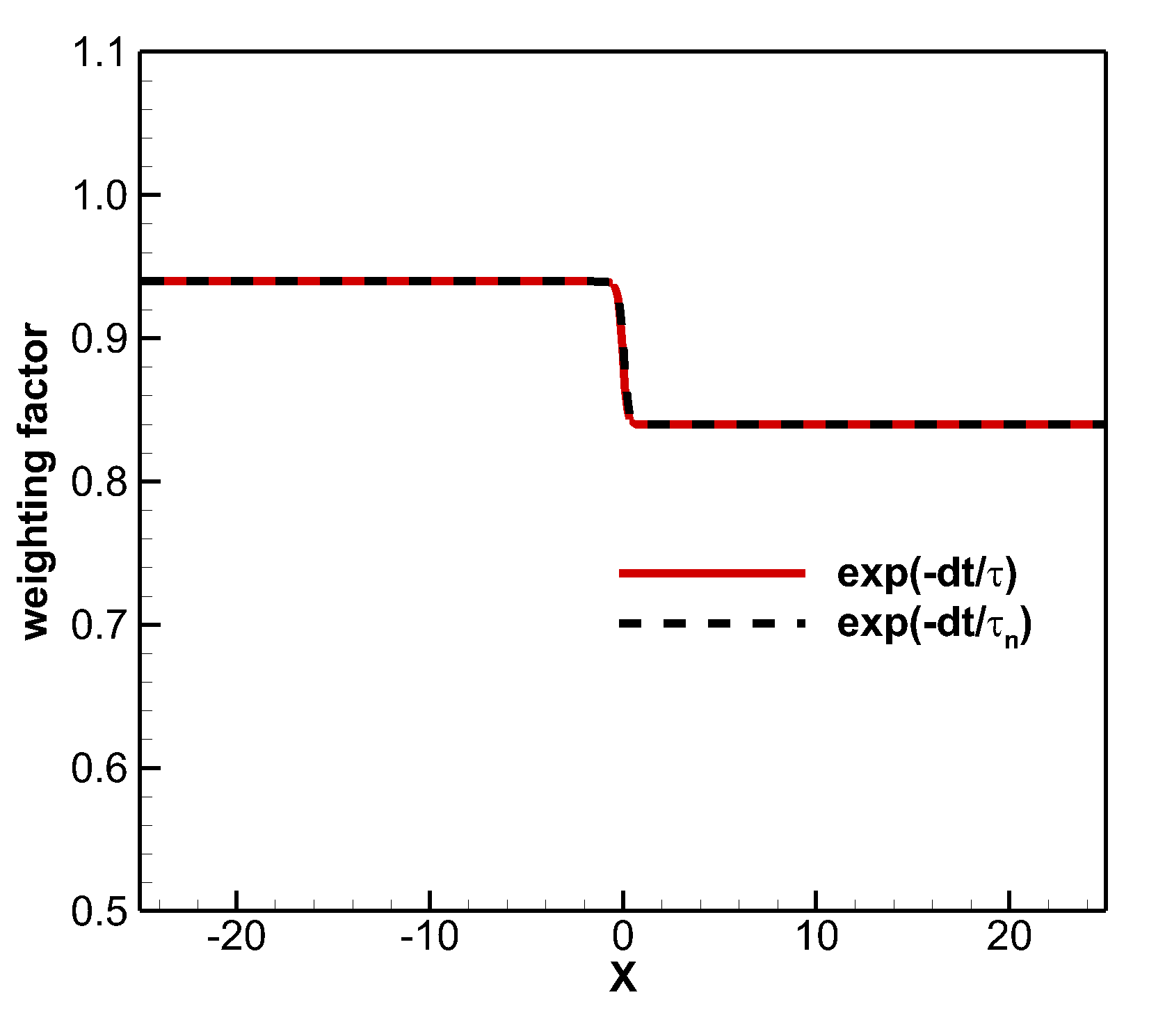}
 \label{shock_Kn0.1_factor}
 }
 \subfigure[wall-clock time]{
	\includegraphics[width=2.2in]{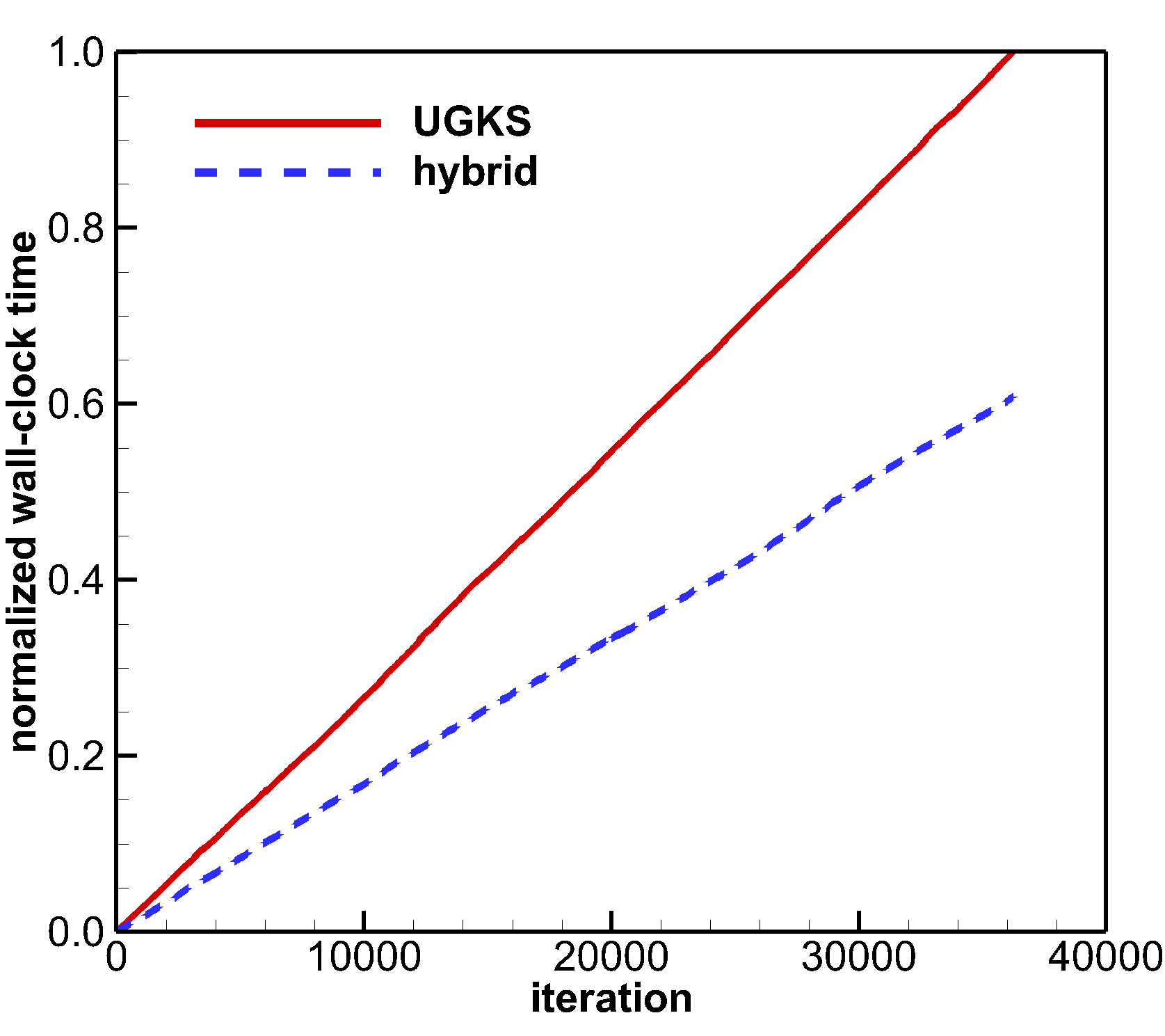}
 \label{shock_Kn0.1_time}
 }
	\caption{The axial distribution of the weighting factors and comparison of computational cost for the shock-structure case at $Kn=0.1$}
\end{figure}

\begin{figure}[h!]
\centering
\subfigure[$\rho$]{
	\includegraphics[width=2.2in]{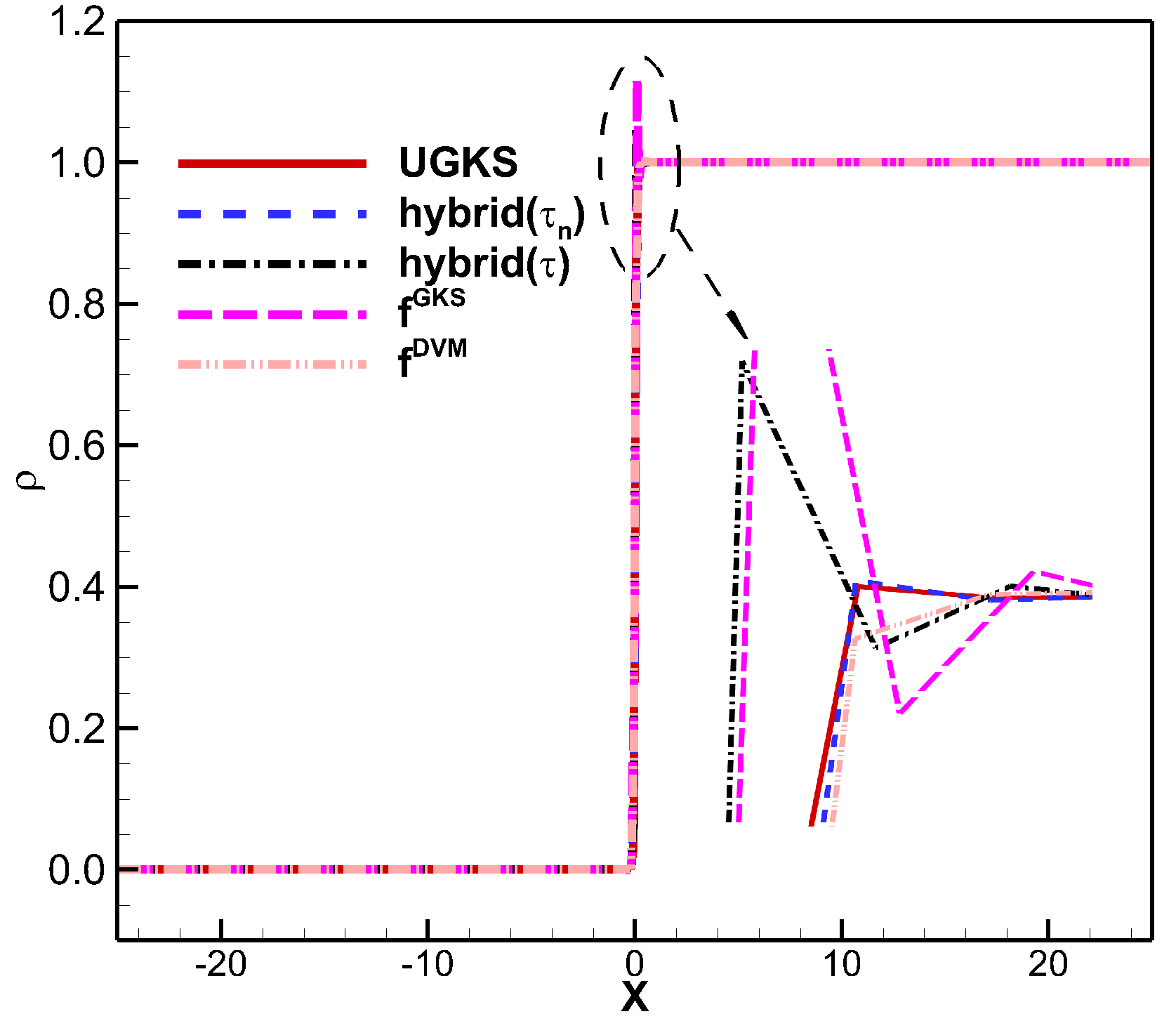}
 \label{shock_Kn0.001_rho}
 }
 \subfigure[T]{
	\includegraphics[width=2.2in]{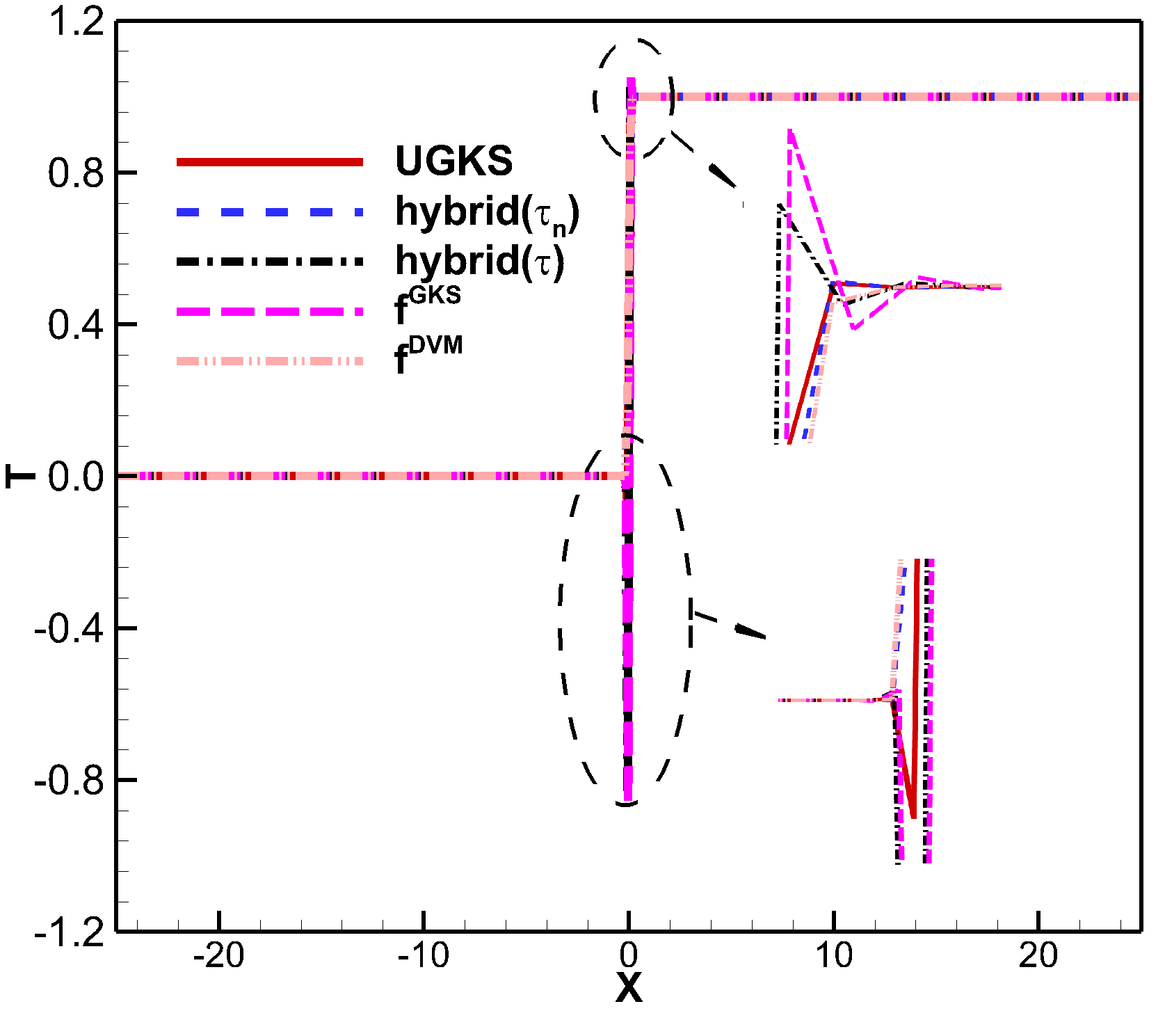}
 \label{shock_Kn0.001_T}
 }
	\caption{The axial distribution of density and temperature for the shock-structure case at $Kn=0.001$}
\end{figure}

Finally, the Knudsen number is further decreased to 0.001, corresponding to the continuum flow regime. Figures~\ref{shock_Kn0.001_rho} and~\ref{shock_Kn0.001_T} compare the axial distributions of density and temperature obtained with different methods. In the legend, $f^{\mathrm{GKS}}$ denotes the equilibrium part of the GKS flux (see Eq.~\ref{fgks}), and $f^{\mathrm{DVM}}$ represents the DVM flux (see Eq.~\ref{fdvm}), while $hybrid(\tau_n)$ and $hybrid(\tau)$ indicate the use of the numerical collision time $\tau_n$ and the physical collision time $\tau$ as the weighting factors in Eq.~\ref{fhybrid}, respectively.

\begin{figure}[h!]
\centering
\subfigure[weighting factor]{
	\includegraphics[width=2.2in]{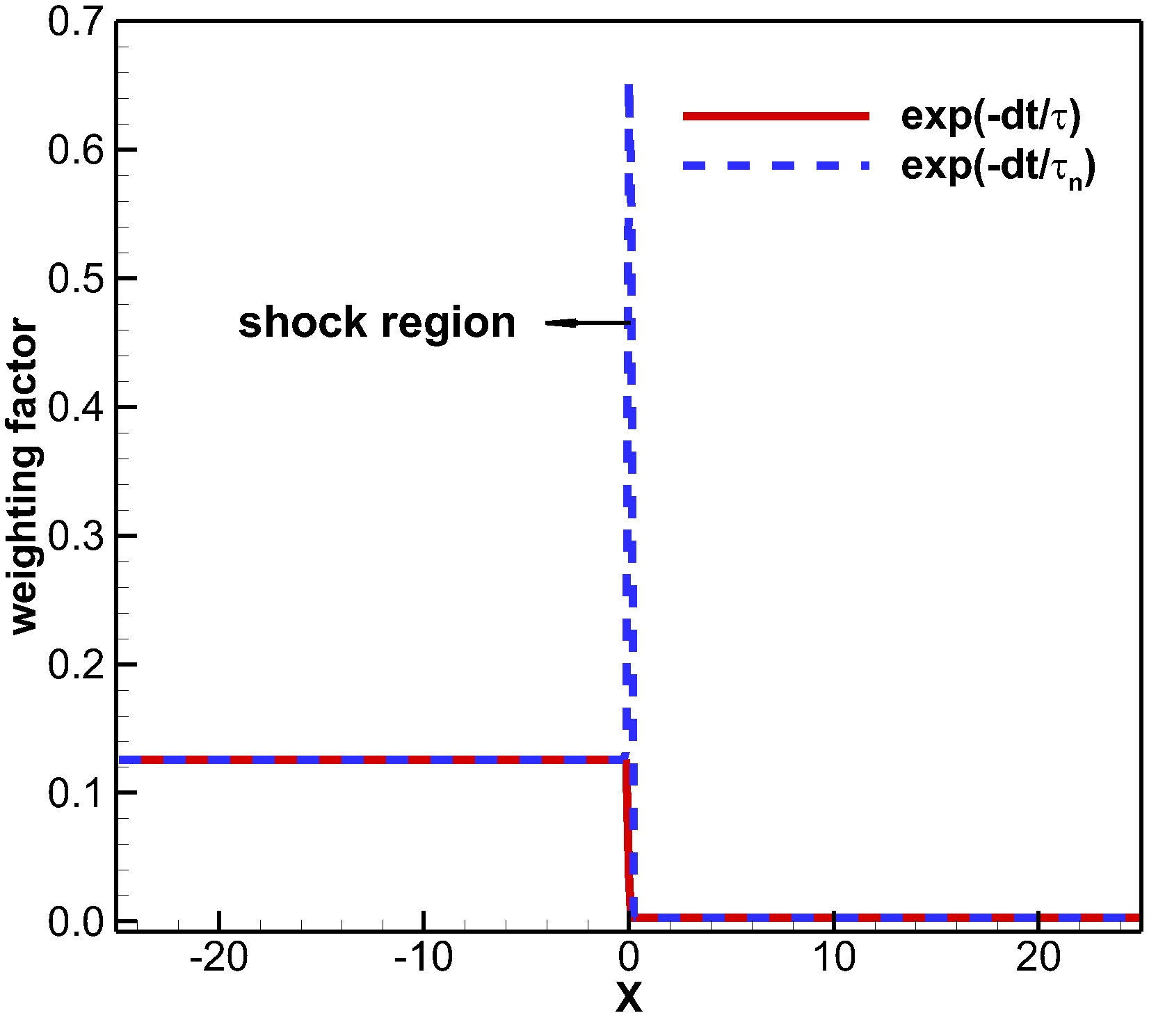}
 \label{shock_Kn0.001_factor}
 }
 \subfigure[wall-clock time]{
	\includegraphics[width=2.2in]{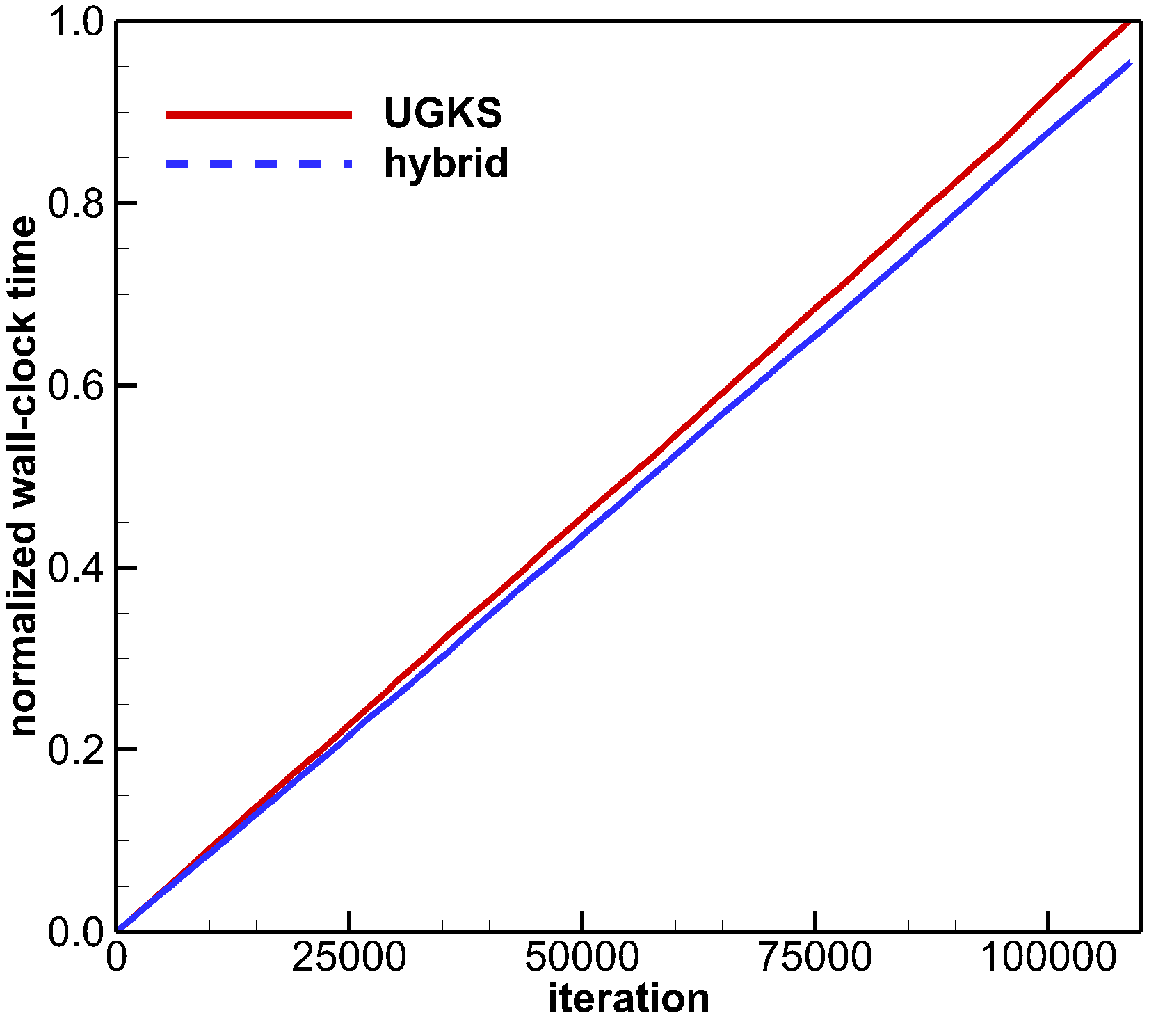}
 \label{shock_Kn0.001_time}
 }
	\caption{The axial distribution of the weighting factors and comparison of computational cost for the shock-structure case at $Kn=0.001$}
\end{figure}

For both density and temperature, pronounced oscillations are observed in the shock region for the methods based on $f^{\mathrm{GKS}}$ and $hybrid(\tau)$. In the continuum regime, where $\Delta t \gg \tau$ and $\exp(-\Delta t/\tau)$ approaches zero (see Fig.~\ref{shock_Kn0.001_factor}), the contribution of the DVM flux becomes negligible, resulting in insufficient numerical dissipation. Consequently, the schemes based on $f^{\mathrm{GKS}}$ and $hybrid(\tau)$ are unable to correctly resolve the sharp shock thickness under the current mesh resolution, leading to numerical oscillations near the shock.
In contrast, the advantage of using the numerical collision time $\tau_n$ as the weighting factor lies in its ability to introduce sufficient numerical dissipation in the shock region while maintaining low dissipation in smooth regions, as evidenced in Fig.~\ref{shock_Kn0.001_factor}. This is attributed to the fact that the DVM contributes appreciably to the numerical flux only in shock-wave regions, while the GKS flux continues to dominate in smooth flow regions. As a result, the hybrid approach employing $\tau_n$ successfully captures the shock structure and preserves high accuracy in smooth flow regions. Although the pure DVM solution can also capture the shock wave, small oscillations appear upstream and downstream of the shock, which are mainly induced by the upwind reconstruction. In terms of computational efficiency, the hybrid approach reduces the overall computational cost by approximately 10\%, as shown in Fig.~\ref{shock_Kn0.001_time}. However, compared with the transition and slip regimes, the acceleration is less pronounced in the continuum regime with shocks, since both the GKS and DVM fluxes must be evaluated.

It is also worth noting that mild oscillations are present in the temperature profile predicted by the UGKS near the shock. This behavior can be attributed to the high accuracy and low numerical dissipation of the UGKS. An effective remedy is to reduce the Courant number, which increases the contribution of the upwind-reconstructed distribution function and thus enhances numerical dissipation.

\subsection{Flow Past a Semi-Cylinder}
\begin{figure}[h!]
    \centering
\includegraphics[width=0.7\linewidth]{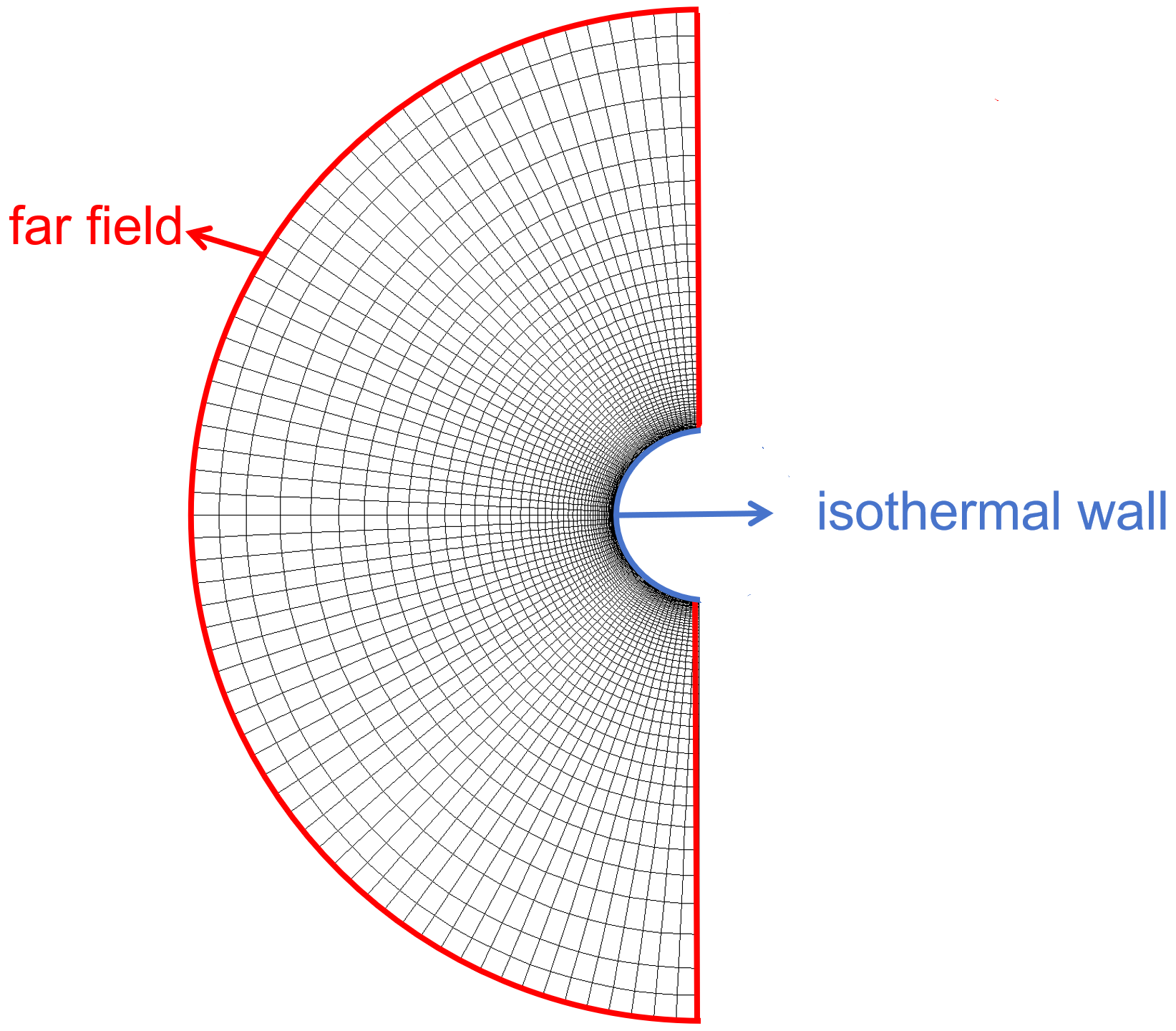}
    \caption{The computational mesh of the flow past a semi-cylinder}
   \label{semi-cylinder_mesh}
\end{figure}

The final test case considers flow past a semi-cylinder. The radius of the semi-cylinder is set to 1.0, and the outer boundary is located at a distance of 6 radius from the body. The free-stream Mach number is set to 5.0, and the isothermal boundary condition is employed to deal with the solid wall. Three Knudsen numbers, namely $Kn$=0.1, 0.01, and 0.001 are investigated. The computations are performed on a structured mesh with a resolution of 71$\times$50, as illustrated in Fig.~\ref{semi-cylinder_mesh}. The velocity space is defined as [-8, 8]$\times$[-8, 8], and is uniformly discretized using 96 velocity points for all Knudsen numbers considered. 

\begin{figure}[h!]
\centering
\subfigure[$Kn=0.1$]{
	\includegraphics[width=4.0in]{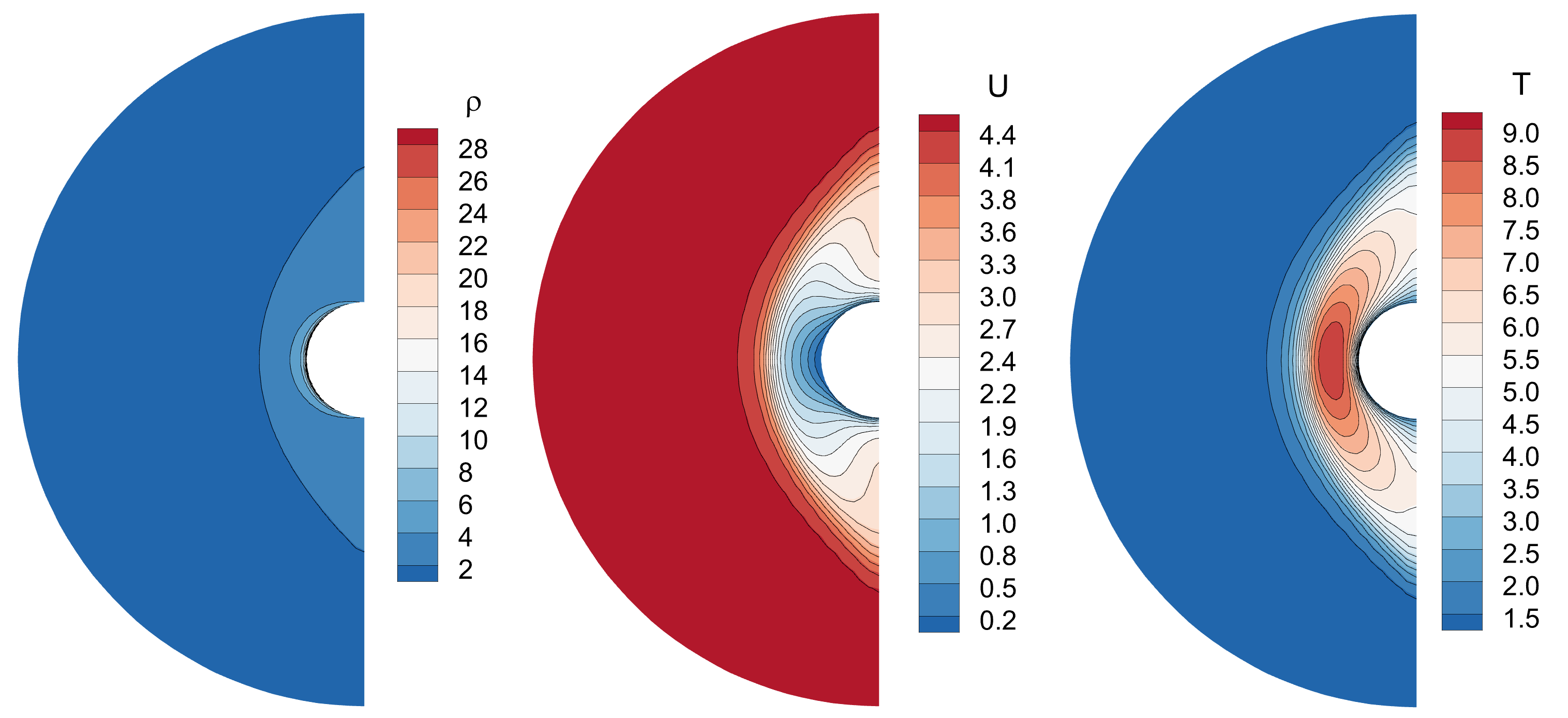}
 \label{semi-cylinder_Kn0.1-contours}
 }
 \subfigure[$Kn=0.01$]{
	\includegraphics[width=4.0in]{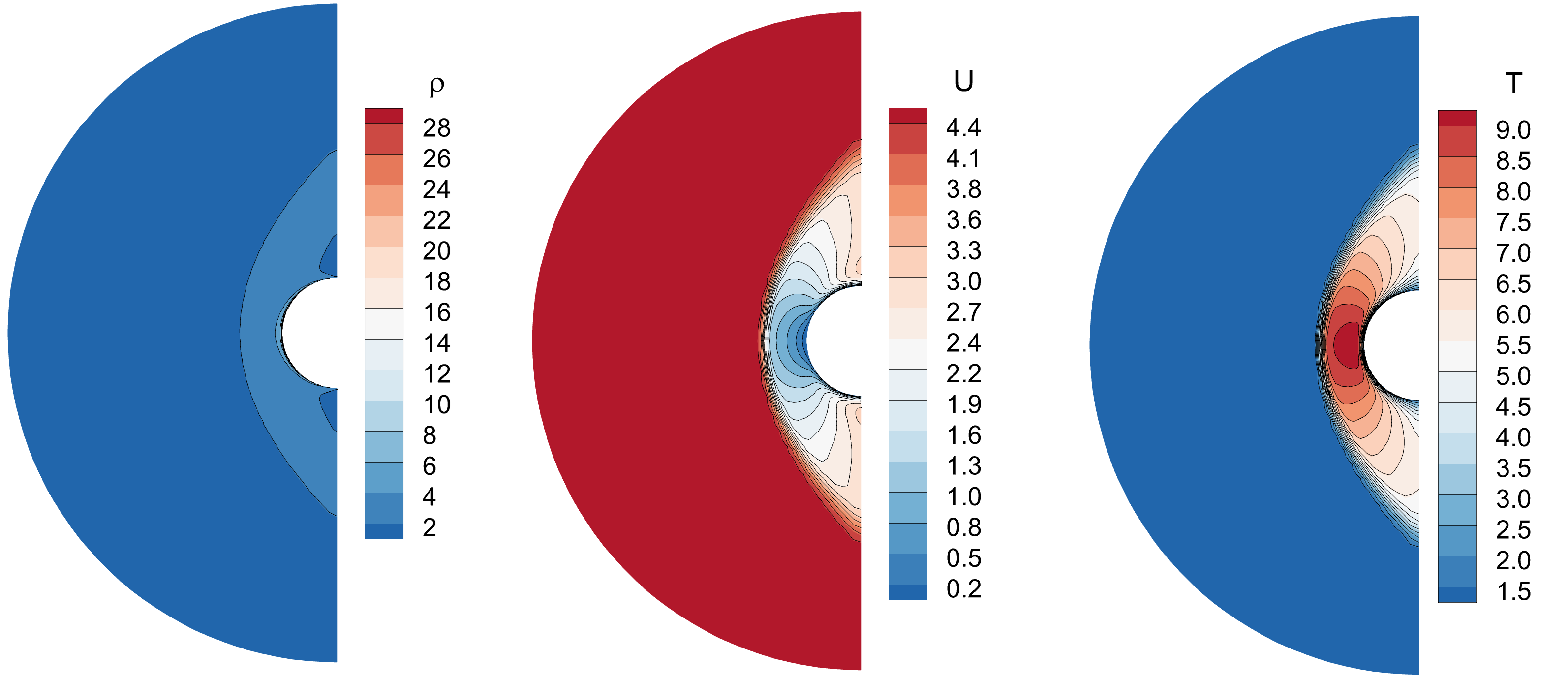}
 \label{semi_cylinder-Kn0.01-contours}
 }
 \subfigure[$Kn=0.001$]{
	\includegraphics[width=4.0in]{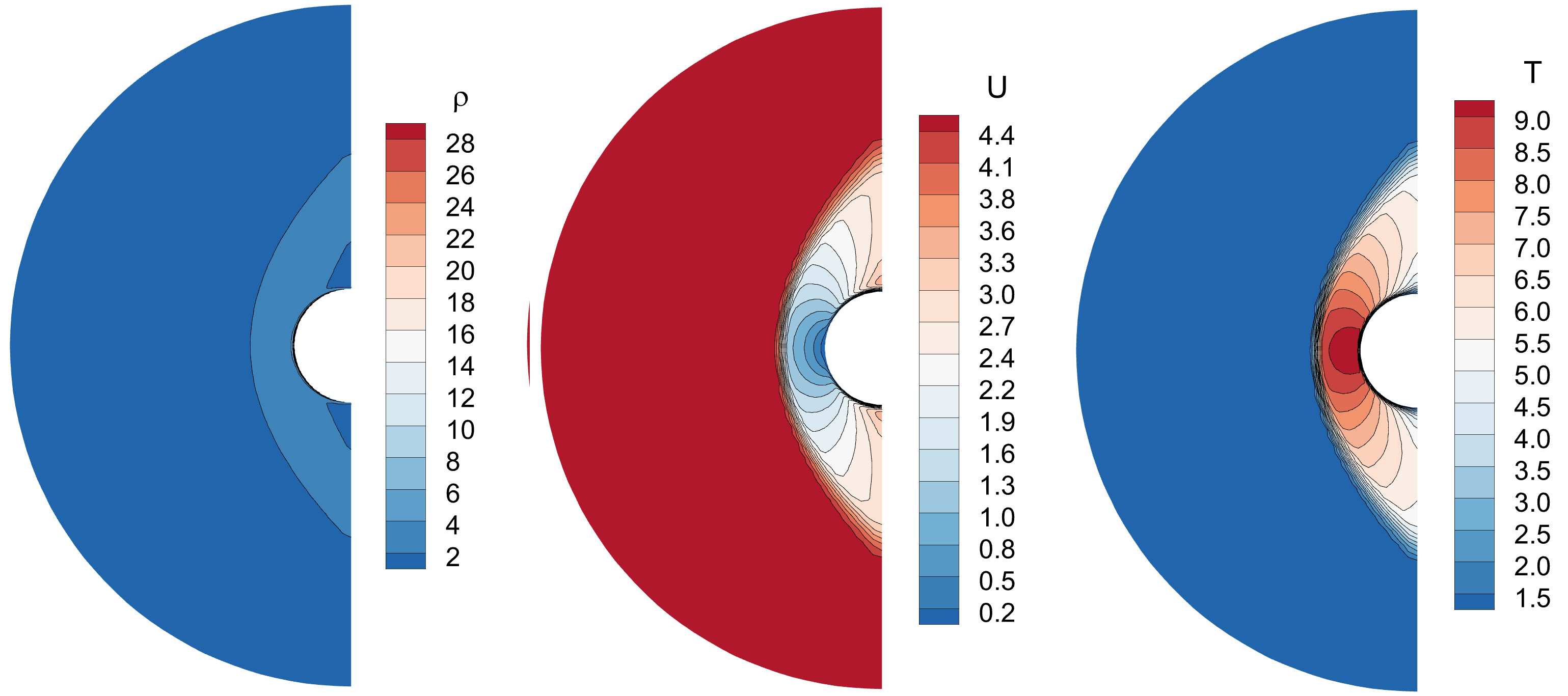}
 \label{semi_cylinder-Kn0.001-contours}
 }
	\caption{Comparison of density, axial velocity, temperature contours at three different Knudsen numbers (the colored backgrounds: UGKS, and the solid lines: the hybrid approach)}
\end{figure}
\begin{figure}[h!]
\centering
\subfigure[$Kn=0.1$]{
	\includegraphics[width=4.5in]{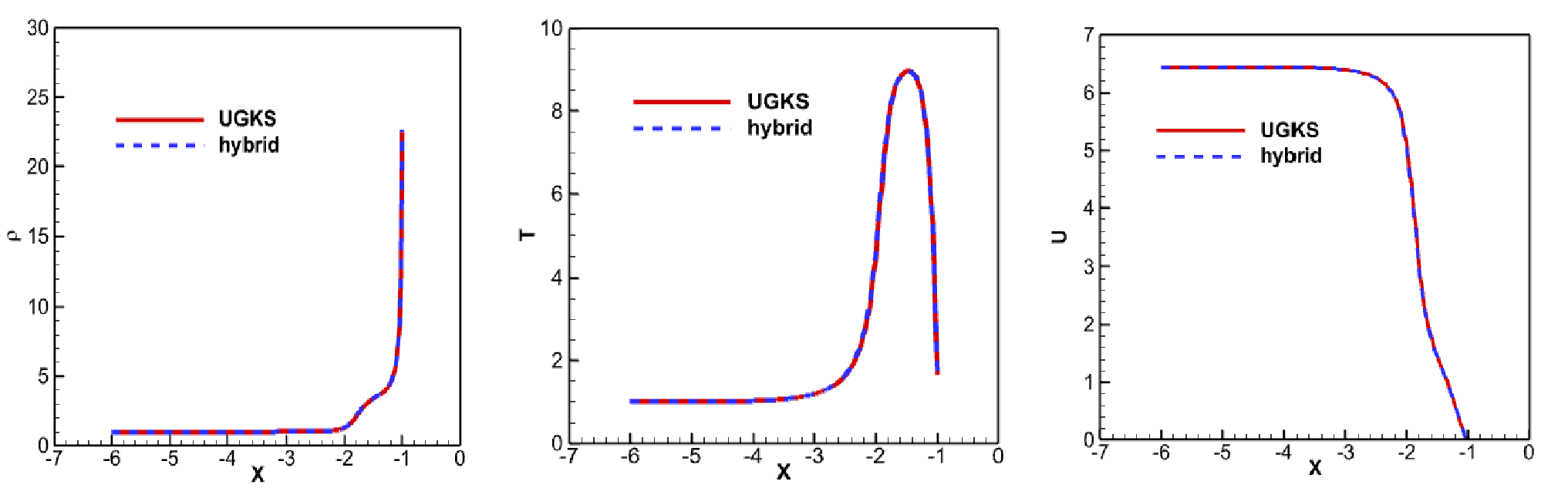}
 \label{semi-cylinder_Kn0.1}
 }
 \subfigure[$Kn=0.01$]{
	\includegraphics[width=4.5in]{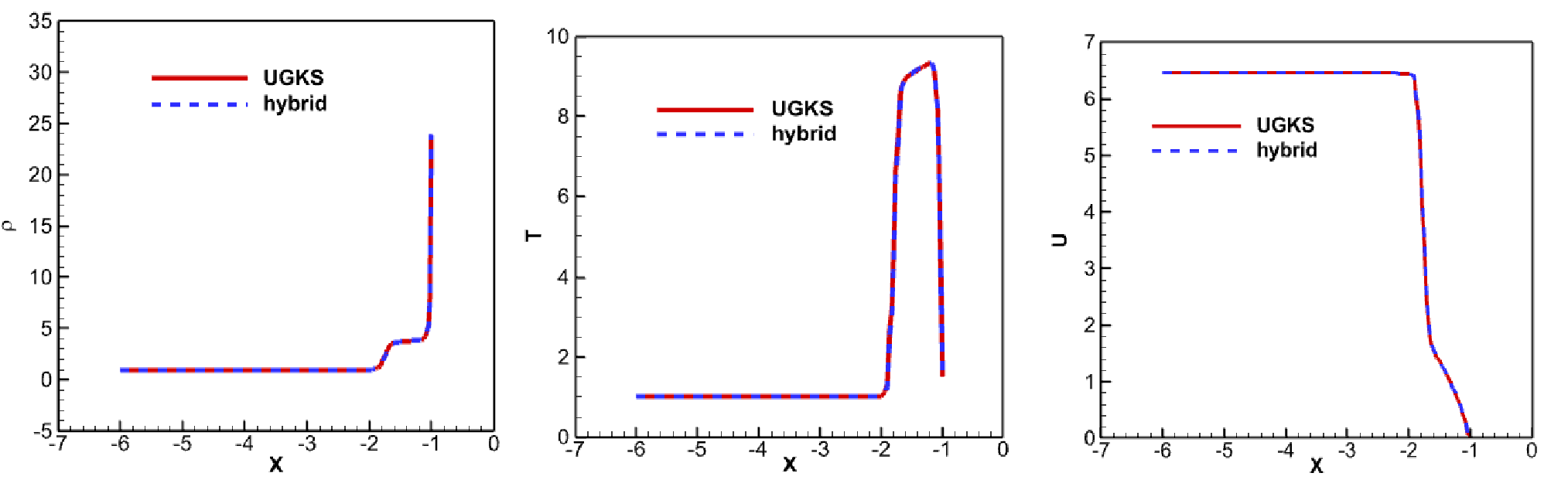}
 \label{semi-cylinder-Kn0.01}
 }
 \subfigure[$Kn=0.001$]{
	\includegraphics[width=4.5in]{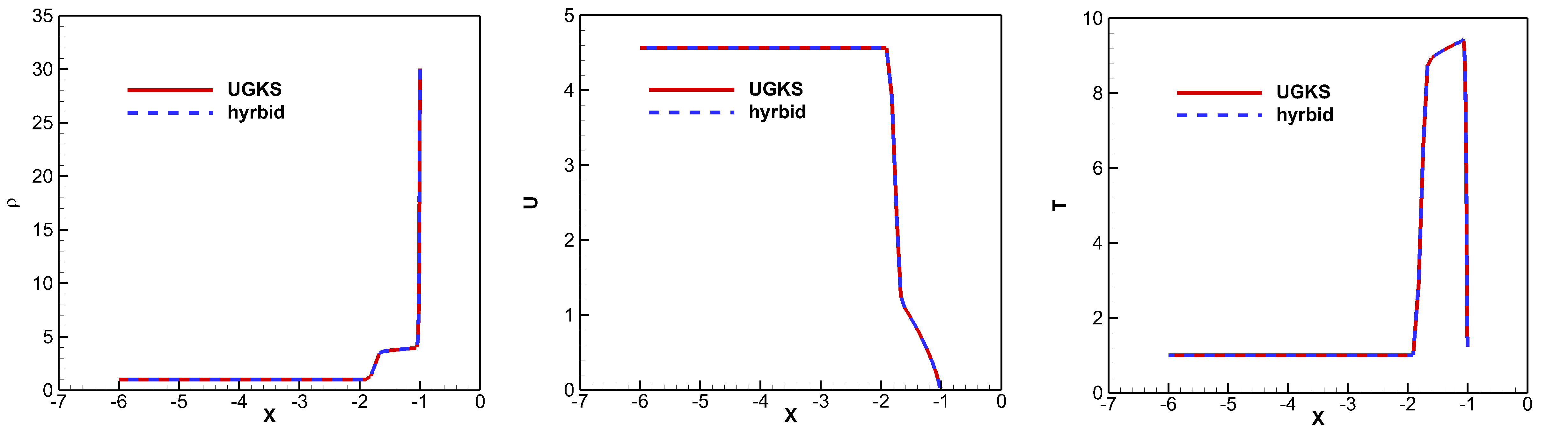}
 \label{semi-cylinder-Kn0.001}
 }
	\caption{Comparison of density, axial velocity, temperature along the axial direction at three different Knudsen numbers}
\end{figure}

Figures~\ref{semi-cylinder_Kn0.1-contours}, \ref{semi_cylinder-Kn0.01-contours}, and \ref{semi_cylinder-Kn0.001-contours} present comparisons of the density, axial velocity, and temperature contours for the three Knudsen numbers considered. In these figures, the colored backgrounds denote the solutions obtained from the UGKS, while the solid lines represent the results from the hybrid approach. Excellent agreement between the UGKS and hybrid solutions is observed across all flow regimes.
Figures~\ref{semi-cylinder_Kn0.1}, \ref{semi-cylinder-Kn0.01}, and \ref{semi-cylinder-Kn0.001} further provide quantitative comparisons of the axial profiles of density, axial velocity, and temperature obtained using different methods. For all three Knudsen numbers, the hybrid approach shows very good agreement with the UGKS solutions.

\begin{figure}[h!]
\centering
\subfigure[residual]{
	\includegraphics[width=2.2in]{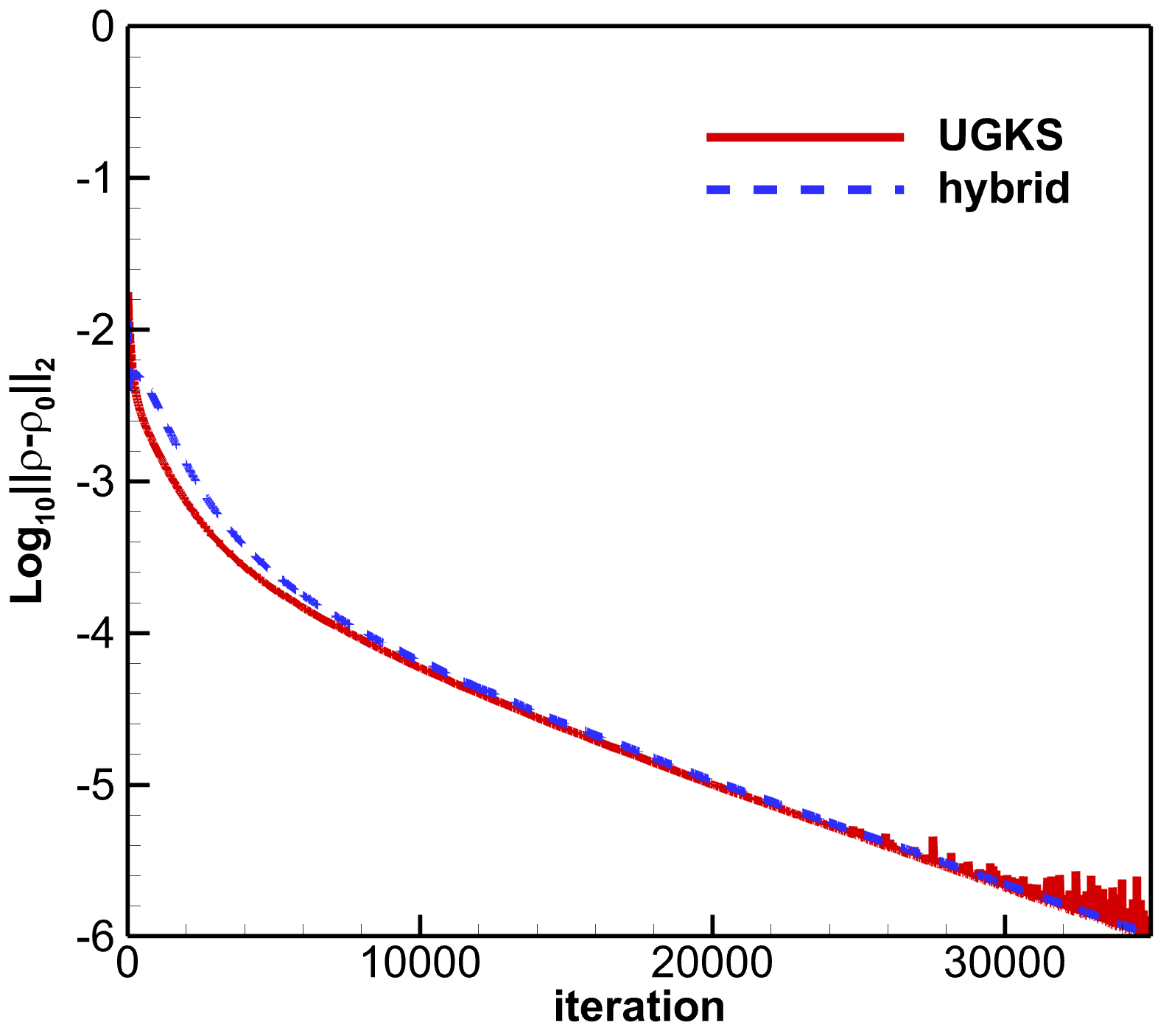}
 \label{semi-cylinder-res}
 }
 \subfigure[wall-clock time]{
	\includegraphics[width=2.2in]{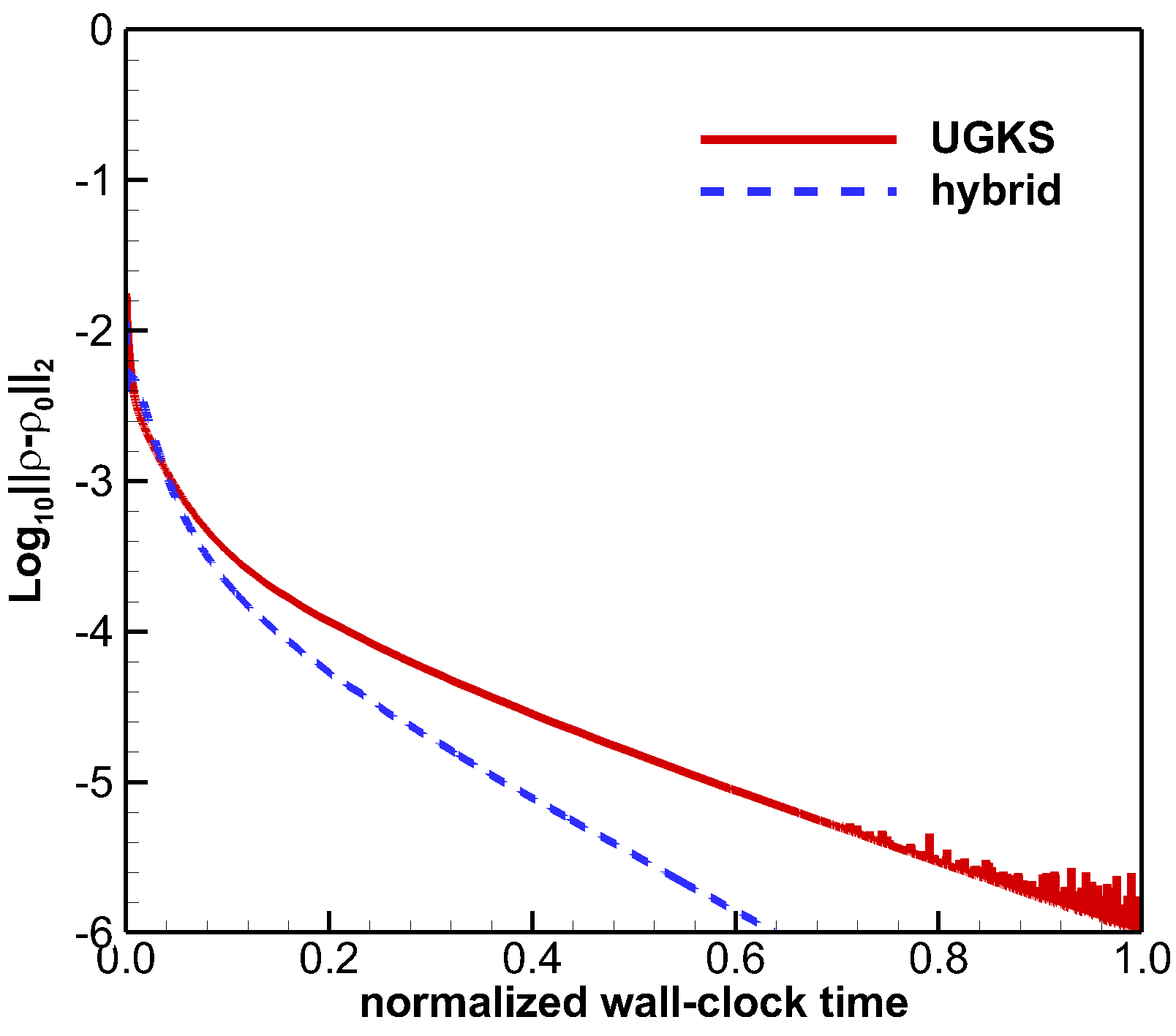}
 \label{semi-cylinder-time}
 }
	\caption{Comparison of the residual convergence histories and wall-clock time between the UGKS and hybrid approach at $Kn=0.01$}
\end{figure}

As demonstrated in Figs.~\ref{semi-cylinder-res} and~\ref{semi-cylinder-time}, the hybrid approach maintains nearly the same residual convergence behavior as the UGKS, but substantially improves efficiency by reducing the computational cost by approximately 1.6 times.

\section{Conclusion}\label{conclusion}

In this work, an efficient hybrid approach is developed for the simulation of steady flows over the entire range of Knudsen numbers. The method couples the equilibrium distribution function from the GKS with the upwind‑reconstructed non‑equilibrium distribution function from the DVM, where a numerical collision time is introduced to balance the two components and ensure accurate shock capturing in the continuum regime. An adaptive strategy based on the prescribed Knudsen and the maximum Mach numbers is further proposed to dynamically switch between modeling components, significantly improving computational efficiency while maintaining accuracy. Asymptotic‑preserving analysis confirms that the proposed method naturally recovers the Navier–Stokes limit in the continuum regime and the free‑molecular solution in the rarefied limit.
Numerical experiments, including flat‑plate boundary layers, lid‑driven cavity flows, shock structures, and flow past a semi-cylinder, demonstrate the robustness and accuracy of the proposed approach across flow regimes ranging from rarefied to continuum. The adaptive strategy reduces the computational cost by approximately one order of magnitude compared with the non‑adaptive hybrid approach. In the rarefied flow regimes, the hybrid method achieves about a two‑fold reduction in wall‑clock time relative to the UGKS. Although the speedup is reduced in continuum flows involving shock waves, a computational cost reduction of approximately 10\% is still achieved. Overall, the proposed hybrid approach exhibits excellent agreement with UGKS solutions over the entire Knudsen number range while delivering substantial computational savings.

 \section*{Acknowledgments}
The authors would like to express sincere gratitude to Mr. Junzhe Cao, Dr. Yue Zhang, and Mr. Hongyu Liu for their valuable suggestions, which have help to further improve the quality of this paper. 
This work was supported by National Natural Science Foundation of China (92371107), Hong Kong Research Grant Council (16208324), and ITF project	PRP/083/24FX.

\bibliographystyle{elsarticle-num} 
\bibliography{main}

\end{document}